\documentclass[aps,pre,twocolumn,longbibliography,showpacs,superscriptaddress]{revtex4-1}
\usepackage{mathrsfs}
\usepackage{graphicx}
\usepackage{wrapfig}
\usepackage{bm}
\usepackage{amsmath,amssymb}
\usepackage{mwe}   % loads »blindtext« and »graphicx«  %%%%THIS IS AWESOME
\usepackage{subfig}

\usepackage{stackengine}
\usepackage[font=normalsize] {caption}
\usepackage{color}
\usepackage[section]{placeins}
\usepackage[english]{babel} 
\makeatletter
\newcommand*{\rom}[1]{\expandafter\@slowromancap\romannumeral #1@}
\makeatother
\newcommand{\red}{\textcolor{red}}

\begin{document}
\title{Microfluidic pump driven by anisotropic \red{phoresis}}
\author{Zihan Tan}\email{z.tan@fz-juelich.de}% \email{tzihankel@gmail.com}
\affiliation{Theoretical Soft-Matter and Biophysics, Institute of
Complex Systems, Forschungszentrum J\"ulich, 52425 J\"ulich, Germany}
\affiliation{Soft Condensed Matter, Institute of
	Complex Systems, Forschungszentrum J\"ulich, 52425 J\"ulich, Germany}
 \author{Mingcheng Yang}\email{mcyang@iphy.ac.cn}
 \affiliation{Beijing National Laboratory for Condensed Matter Physics and Key Laboratory of Soft Matter Physics, Institute
 of Physics, Chinese Academy of Sciences, Beijing 100190, China}
  \affiliation{University of Chinese Academy of Sciences, Beijing 100049, China}
\author{Marisol Ripoll}\email{m.ripoll@fz-juelich.de}
\affiliation{Theoretical Soft-Matter and Biophysics, Institute of
Complex Systems, Forschungszentrum J\"ulich, 52425 J\"ulich, Germany}
\date{\today}

\begin{abstract}
  Fluid flow along microchannels can be induced by keeping opposite
  walls at different temperatures, and placing elongated tilted
  pillars inside the channel. The driving force for this fluid motion
  arises from the anisotropic thermophoretic effect of the elongated
  pillars that generates a force parallel to the walls, and
  perpendicular to the temperature gradient. The force is not
  determined by the thermophilic or thermophobic character of the
  obstacle surface, but by the geometry and the thermophoretic
  anisotropy of the obstacle. Via mesoscale hydrodynamic simulations,
  we investigate the pumping properties of the device as a function of
  the channel geometry, and pillar surface properties. Applications as
  fluidic mixers, and fluid alternators are also outlined, together
  with the potential use of all these devices to harvest waste heat
  energy. Furthermore, similar devices can be also
  built employing diffusiophoresis or electrophoresis.
\end{abstract}

%\pacs {02.70.-c,47.11.-j,47.57.J-}
% 02.70.-c Computational techniques;simulations
% 47.57.J-  Colloidal Systems
% 47.11.-j Computational methods in fluid dynamics

\maketitle

\section{Introduction}
Guiding the movement of fluid at nano- and micro-scales has become one
of the most challenging goals in the emergent field of
microfluidics~\cite{Whitesides2006,heatengine}. Relevant applications
of an efficient design of microfluid pumping are related with drug
delivery~\cite{nisar2008,herrlich2012}, biomedical
assays~\cite{wang2018,Amirouche2009} and cell
culturing~\cite{Byun2013}. This requires the capability to use minimal
quantities of energy to carry fluid with high resolution and
sensitivity. In microfluidic devices, fluids are typically
transported, separated, or processed along microchannels of different
compositions and geometries. The generation of net fluid flows is
frequently achieved by applying external mechanical forces with
coupled inlet and outlet
systems~\cite{RevModPhys.77.977,stone2004engineering,darhuber2005principles}.
The efficiency of such driving mechanisms importantly decreases with
miniaturization because of the huge increase in hydrodynamic
resistance that comes with downsizing~\cite{Dauparas2018}. Moreover,
the fact that in- and outlets rely on external pieces of equipment,
importantly hampers the portability of the devices. A competitive
approach is to induce stresses localized at the boundaries, through
non-mechanical means which are driven typically by local
fields~\cite{Zhou2016ChemistryPumps,sebaGeometric,16ratchet,PaxtonMotor,Thomas2016}.
This has shown to be especially efficient for miniaturizing fluidic
pumps, due to the intrinsic large surface to volume ratio.

Phoretic or the related osmotic properties of solid materials in a fluid solution
constitute an attractive option to induce stresses close to confining
walls.  Phoresis refers to the directed drift motion of a suspended
particle induced due to an inhomogeneous surrounding~\cite{Anderson,Moran2017}, which can be,
for example, a temperature gradient (thermophoresis)~\cite{Alois2010,PiazzaReview,frenkel2017,Fu2017}, a concentration
gradient (diffusiophoresis)~\cite{Derjaguin,brady2011,Howse2007} or an electric potential gradient
(electrophoresis)~\cite{Saville1977,Shendruk2012,PhysRevLett.109.098302}. Conversely, the gradient can generate the motion of
a fluid at a fixed solid-fluid interface, which is usually referred to
as phoretic osmosis, and in the case of a fluid-fluid interface, which
is known as phoretic capillary~\cite{Anderson}.  Catalytic
surfaces and related chemical gradients have shown a large potential
in microfluidic applications~\cite{Paxton2006,sengupta2014,Ortiz-Rivera2016,das2017}, while thermal gradients are relatively
less exploited. Thermal gradient-driven motion has though promising
prospectives since it works equally well in charged and neutral
solutions, and it is pollution-free due to the absence of surfactants
or chemical fuels, which enables the way to bio-compatible
applications~\cite{duhr2004,vigolo2010,tsuji2017}. Furthermore,
thermal gradient-driven motion allows optical microscale operations
with optical heating which is the basic principle of the emerging
field of optofluidics~\cite{baigl12,namura2017quasi}. So far, existing
phoretic fluidics rely on intricate differentiated
compositions~\cite{kline2005,Jiang2010,PhysRevLett.105.174501} or ratchet
geometries of channel
walls~\cite{yang14gear,Yang15cgear,16ratchet,leonardo15gear}.
To extend the tunability and functionality of these pumps is therefore
timely and highly desirable.

Anisotropic thermophoresis has been recently described opening a new
avenue for the design of novel and versatile
microdevices~\cite{ZT01,YangHeat}. %%
The anisotropic phoretic effect refers to the different phoretic
response that elongated objects, such as colloids or pillars, have
when aligned with, or perpendicular to the external
gradient. Interestingly, for obstacles with tilted orientation this
anisotropy might translate into a force which would not only be
aligned with the gradient, as it is in the case of traditional phoresis, but
additionally also perpendicular to the gradient.  This mechanism has
been already employed for the design of microturbines which can rotate
unidirectionally~\cite{YangHeat}. Given the related nature of
different phoretic effects, this effect has proved to exist not only
for thermohoresis, but also for diffusiophoresis, this is in for
example in the presence of multicomponent fluids with catalytic
surfaces~\cite{YangMass}. 
% To further exploit applications of
% anisotropic phoresis in microfluidics is certainly very valuable.

In this work, we propose a new class of microfluidic pump based on the
anisotropic phoretic effect. Instead of considering the asymmetries at
the channel walls as in previous devices, we present a micropump
exploiting the tunable properties of the immersed solid obstacles with
thermophoretic anisotropy in the middle of the channel, which could be
engineered for example by means of
lithography~\cite{Tasinkevych16268,Liu9231}. Opposite walls will have
fixed different temperatures, such that the pillars are exposed to a
temperature gradient. The phoretic properties of the pillars surface
will thus generate a fluid flow along the microchannel, perpendicular
to the thermal gradient. The device will have a large versatility due
first to factors intrinsic to the anisotropic pump such as pillars
geometry, number of obstacles per unit length, or obstacles
configuration, the subtle dependence of the phoretic behavior on a
large number of factors such as average temperature or pressure,
presence of salt, or surface coatings among many others. Also very
remarkable will be the case in which the obstacles do not have a fixed
orientation with respect to the walls, but can be externally in situ
controlled, e.g. by laser tweezers on suspended colloids, such that
the flow pattern will become highly adjustable and rich. In addition,
the suspended objects may even be removed from and imported into the
channel without affecting the channel itself. Furthermore, variations
of the microchannel device are shown to work not only as microfluidic
pumps with arbitrarily microchannel length, but also as fluidic
mixers, or generators of alternating flow. Finally, and given the fact
that they work under the effect of external gradients, they all have
potential applications to harvest waste thermal or chemical energy.

%%%%%%%%%%%%%%%SECTION TWO: MODEL AND METHODS%%%%%%%%%%%%%%%%%%%%
\section{Model and mechanism}

\subsection{Simulation setup}

Simulations have been performed by a mesoscale hydrodynamic approach
which combines multiple particle collision dynamics (MPC) and
molecular dynamics (MD)~\cite{kap99,kap00,kapral_review,MPCD}. %%
MPC is a particle-based model in which a coarse grain solvent is
represented by $N$ point particles of mass $m$, characterized by
continuous positions ${\bf r}_i$ and velocities ${\bf v}_i$
($i=1,\ldots,N$), which evolve in two alternating steps.  In the
\textit{streaming step}, particle positions evolve ballistically for a
certain time $h$, which we refer to as the {\em collision time}.  In
the \textit{collision step}, particles are sorted into cubic cells,
with cell size $a$, and each particle rotates its relative velocity to
the cell center of mass velocity by an angle $\alpha$ around a
randomly chosen direction. The collisional operation conserves mass,
linear momentum and energy at the cell level. The cell-grid is
randomly displaced to maintain \textit{Galilean
  invariance}~\cite{ihl01} and enhance the fluid like properties of
the solvent~\cite{ihl03,tuz06,pre05}.  %
Standard simulation units are chosen, $m = k_BT = a = 1$, with $k_BT$
the averaged temperature, which means for example that time is given
in units of $\sqrt{ma^2/(k_BT)}$, which will be typically not
specified.  Usual MPC parameters are used, $h=0.1$, $\alpha=130^o$,
and $\rho=10$ average number of particles per collision box,
corresponding to a kinematic viscosity $\nu=0.87$.  For direct
comparison with real fluids, dimensionless numbers are typically
calculated, such as the Schmidt number $Sc$, or the Prandlt number
$Pr$, which for the specified parameters are $Sc=17$ and $Pr=5.8$
respectively. In this respect $Pr$ shows to be very close to most
relevant fluids such as water, although $Sc$ is clearly smaller. While
the Prandlt number indicates that the transport of heat and momentum
are properly separated in our case, the Schmidt number refers to the
separation of mass and momentum, which in our case are less separated
than in most fluids. This lower value might be an issue when
quantitatively mapping to real units, although is absolutely not a
problem to reproduce a fluid with correct liquid-like
dynamics~\cite{epl04,pre05}. Solid channel walls with stick boundary
conditions are realized by applying the bounce-back rule to the
solvent particles when reaching the walls, and the temperature
gradient is implemented by thermostatting thin solvent layers close to
the walls with low $T_c$ and high $T_h$
temperatures~\cite{luese12jcp,yang13flow}.  The employed default
temperature at the no-slip walls are $T_h=1.2$ and $T_c=0.8$ in MPC
units. These could be converted to real units by considering for
example an average temperature equal to $T=300K$, corresponding to
very high temperature differences, although, as later discussed, this
could be compensated with other unit mismatches.

\begin{figure}[t]
	\centering
					\subfloat{		
		%\centering
		\begin{picture}(100,100)
		\put(-40,-0){\includegraphics[width=0.35\textwidth,trim={0 0 0 0},clip]{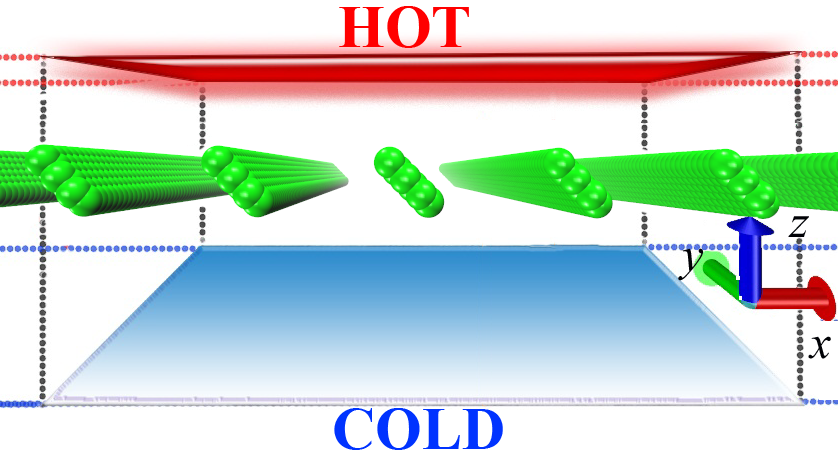}}
		\put(-52.0,88){(a)}
		\end{picture}
		\label{fig:3dscheme}
	}
\quad\quad\quad\quad
	%\vspace{0.5cm}
				\subfloat{		
		%\centering
		\begin{picture}(100,100)
		\put(-40,5){\includegraphics[width=0.35\textwidth,trim={0 0 0 0},clip]{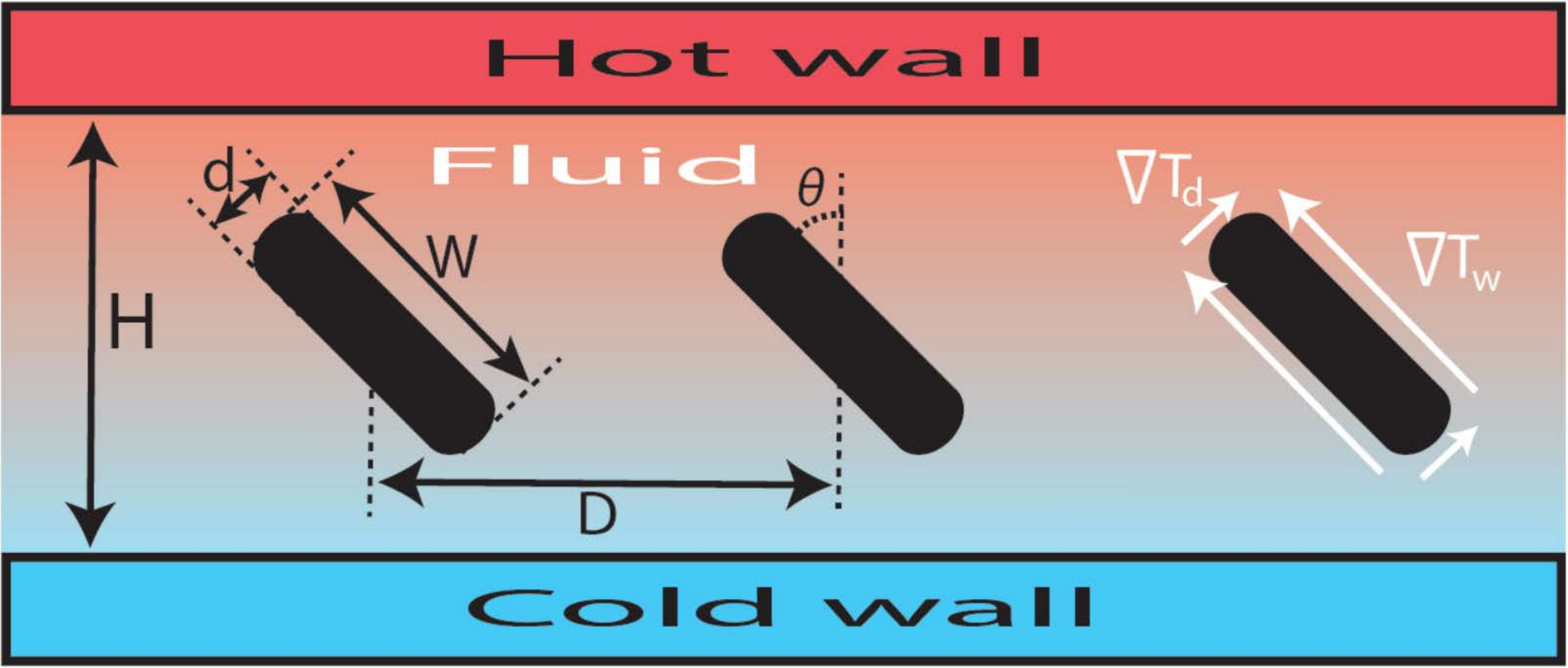}}
		\put(-52,70){\textcolor{black}{(b)}}
		\end{picture}
		\label{fig:2dscheme}
	}
	\vspace{0.3cm}
			\subfloat{
		%\raggedleft
		\def\stackalignment{l}
		\topinset{(c)} {\includegraphics[width=0.38\textwidth]{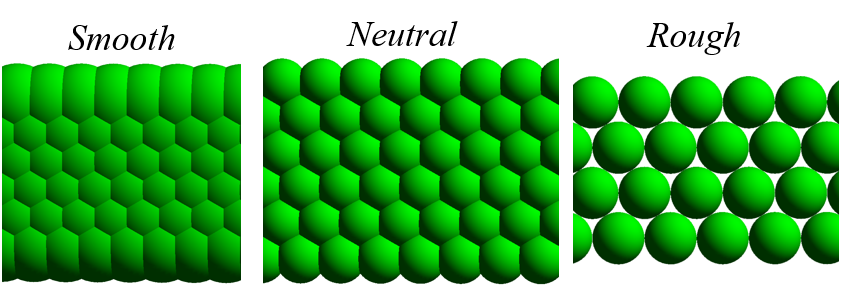}}{0.05in}{-0.1in}  
		\label{fig:obstacle}
	}
	\caption {(a)~Sketch of the anisotropic micropump, composed by
          two planar walls at fixed different temperatures, with solid
          elongated tilted in between obstacles. (b)~Cross-section
          sketch of the relevant device dimensions. The channel width
          is $H$, each pillar has a length $W$ and thickness $d$,
          and multiple obstacles are separated by a distance $D$. The
          temperature gradients along the long and short axes are
          $\nabla T_d$ and $\nabla T_W$. (c)~Detailed structure of
          pillars with rugosities $l/d=0.4$, $0.8$, and $1$,
          referred to as smooth, neutral, and rough.}
	\label{schemes}
\end{figure}

The thermophoretic micropump, sketched in Fig.~\ref{schemes}, consists
of a microchannel of width of $H$ with immersed solid elongated
obstacles (pillars) tilted at an angle $\theta$, typically
$\theta=45^o$. Opposite walls have fixed different temperatures, such
that the pillars feel a temperature gradient $\nabla T$, which will
induce an osmotic flow. %%
As indicated in Fig.~\ref{schemes}\subref{fig:obstacle}, pillars are
modeled by one layer of spherical beads of the diameter $d$, fixed in
this work to $d=2.5a$, placed with anchored positions on the nodes of
a triangular lattice. The separation between beads determines the
pillar surface rugosity, which is characterized by the ratio
between the typical lattice length $l$, and the bead diameter $d$. In
this way small values, such as $l/d=0.4$ are refereed to as smooth
surfaces, large values, such as $l/d=1$ are referred to as rough. %
In order to modify the surface thermophoretic properties rugosity is
here modified similar to the rod-like colloid case.  The rough case
corresponds then to porous like material where fluid can cross through
the pillars interstices. %%
The pillar thickness is for all rugosities, the particle
diameter $d$, and the length is denoted by $W$, which is typically not
an integer given the varying values of the rugosity. Periodic boundary
conditions (PBC) are considered in the two directions perpendicular to
the walls. Pillars are placed with the long axis parallel to the
  walls, this is in the y direction as depicted in Fig.~1a. This,
  together with PBC, provides effective extended pillars. Finally, the
  pillars are equidistantly placed in the $x$ direction, with
  separation $D$. Unless otherwise stated, we perform simulations
with $l/d=0.4$, $W=2.93d$, $D=H=12d$, and with one obstacle in a
cuboid box $(L_x, L_y, H)=(30, 20, 30)$.  The beads positions are
invariable with time, and the interaction of the solvent particles
with the pillar beads are modelled by Mie potential~\cite{miexx}
(namely, Lennard-Jones type)
\begin{equation}
	U(r) =4\epsilon\left[\left(\frac{d}{2r}\right)^{2n}
	-\left(\frac{d}{2r}\right)^{n}\right]+c, \quad r\leq r_c. 
	\label{lj}
\end{equation}
Here $r$ is the distance between the bead center and the fluid
particle, $\epsilon$ is the potential intensity, chosen as
$\epsilon=k_BT$, and $n$ is a positive integer describing the
potential stiffness. By considering $c=0$ or $c=\epsilon$ together
with the suitable cutoff distance $r_c$, the potential is adjusted to
be attractive or purely repulsive~\cite{luese12jpcm}, which will be
respectively denoted as $an$ or $rn$.  This means for example, that
the $r6$ potential with $n=6$, $c=\epsilon$, and $r_c=2^{1/6}d$ is the
well-known WCA \mbox{(Weeks-Anderson-Chandler)} repulsive
potential~\cite{wca1971}.

\subsection{Mechanism: anisotropic thermophoresis}
\label{sec:mech}

The interaction of a colloidal particle with a solvent with a
non-homogeneous temperature results in a driving thermophoretic force
$\textbf{F}_T$~\cite{yang12forces,yang12drift} which causes the
particle migration of non-fixed particles (thermophoresis), or the
motion of the surrounding fluid for immobilized objects
(thermoosmosis).  This force is known to be directly proportional to
the temperature gradient,
\begin{equation}
	\textbf{F}_T=-\Lambda_T \cdot k_B\nabla T,
	\label{eq_sphere}
\end{equation}
with $k_B$ the Boltzmann constant, and $\Lambda_T$ the thermodiffusion
tensor, which is a material dependent property determining the force
direction and strength. In the case of spherical particles,
$\Lambda_T$ is a constant factor, the so-called {\em thermodiffusion
  factor}, $\alpha_T$; while other particle shapes might have more
complex dependencies. This deviation from the spherical constant
behavior has been defined as {\em anisotropic
  thermophoresis}~\cite{ZT01}. In the case of elongated particles, two
independent coefficients are expected to be enough to determine the
thermophoretic properties. These are $\alpha_{T,\|}$ and
$\alpha_{T,\perp}$, the thermophoretic factors characterizing a rod
with the long axis aligned with the temperature gradient, or
perpendicular to it. The difference between these two factors defines
the {\em thermophoretic anisotropy factor}
\begin{equation} \label{chi}
\chi_T=\alpha_{T,\perp}-\alpha_{T,\|}.
\end{equation}

Interestingly, this means that an elongated particle with $\chi_{T}
\neq 0$ fixed an angle $\theta$ with respect to $\nabla T$ will feel 
a force not only in the gradient direction, but also perpendicular to it~\cite{ZT01}
\begin{equation}
{\bf F}_{T,x}=-\chi_T \sin\theta \cos\theta k_{B}|\nabla T|{\bf e}_{x},\label{perp} 
\end{equation}
where $x$ refers to the direction perpendicular to $\nabla T$, and
${\bf e}_{x}$ the corresponding unit vector. %%
By convention, the sign of the thermodiffusion factor $\alpha_T$ is
positive when colloids drift toward cold areas, this is for
\textit{thermophobic} behavior; while $\alpha_T$ is negative for
colloids drifting towards warmer areas, this is \textit{thermophilic}
behavior. Previous simulations showed that the use
of attractive solvent-colloid potential translates into a thermophobic
colloid drift, while repulsive potentials show a thermophobic
behaviour. This driven mechanism of the model is essentially related
to the relation between the local pressure gradient and the sign of
the applied potential~\cite{ZTphd,Burelbach2018}. %%
The direction of the perpendicular force $F_{T,x}$ in Eq.~(\ref{perp})
is determined by the sign of $\chi_T$ which is in principle
independent of the thermophobic or thermophilic character of the surface. 

In the microchannel configuration of Fig.~\ref{schemes}, the fixed
solid pillars endure thermoosmotic % forces and
flows of the fluid around them. If the obstacles are elongated
structures ($W\neq d$), tilted with respect to the channel walls and
therefore to the temperature gradient ($0^\circ<\theta<90^\circ$), a
net fluid flux will be generated parallel to the channel walls; as can
be seen in Fig.~\ref{fig:sch_flow}.  The resulting net flux density
$\hat{J}$ can be defined as the particle flux per unit volume,
\begin{equation}
\hat{J}(x)
=\frac{1}{H}\int^H_0 \rho(x,z) v_x(x,z)dz,
\label{flux}
\end{equation}
where $\rho(x,z)$ is the average particle number, and $v_x(x,z)$ the
particle velocity of the fluid at the position $(x,z)$. Note that the
fluid velocity is not fully considered, but just the velocity
component parallel to the walls, since this is  the only
one that contributes to the net flux. The system symmetry allows us to
disregard the system dependence along the obstacle length, which in
Fig.~\ref{schemes}\subref{fig:3dscheme} corresponds to the
$y$-direction. %%
In order to provide a prediction to the corresponding averaged density
flux, it is important to note that the normalized number density can
be considered proportional to $\rho_0$, the averaged number density;
while the fluid velocity will be determined by the perpendicular
thermophoretic force ${\bf F}_{T,x}$ in Eq.~(\ref{perp}), together
with an effective friction coefficient proportional to the fluid
viscosity $\eta$.  The averaged density flux will be then determined
by,
\begin{equation}
\widetilde{J} =  G(W,D,H) \frac{\rho_0}{\eta} \chi_T  k_B \vert \nabla T \vert
\label{eq-jxa}
\end{equation}
where a fixed angle inclination has already been accounted for, for
example its optimal value $\theta=\pi/4$, and $G(W,D,H)$ is a function
of the microchannel and pillars geometry with inverse of length
dimensions. %%
Expression~(\ref{eq-jxa}) indicates that the intensity of the flux
will be determined by several parameters, but the direction will be
remarkably only determined by the anisotropic thermophoretic factor
$\chi_T$.

\section{Flow pattern and flow flux}
\begin{figure*}
	\centering
\includegraphics[height=0.27\textwidth]{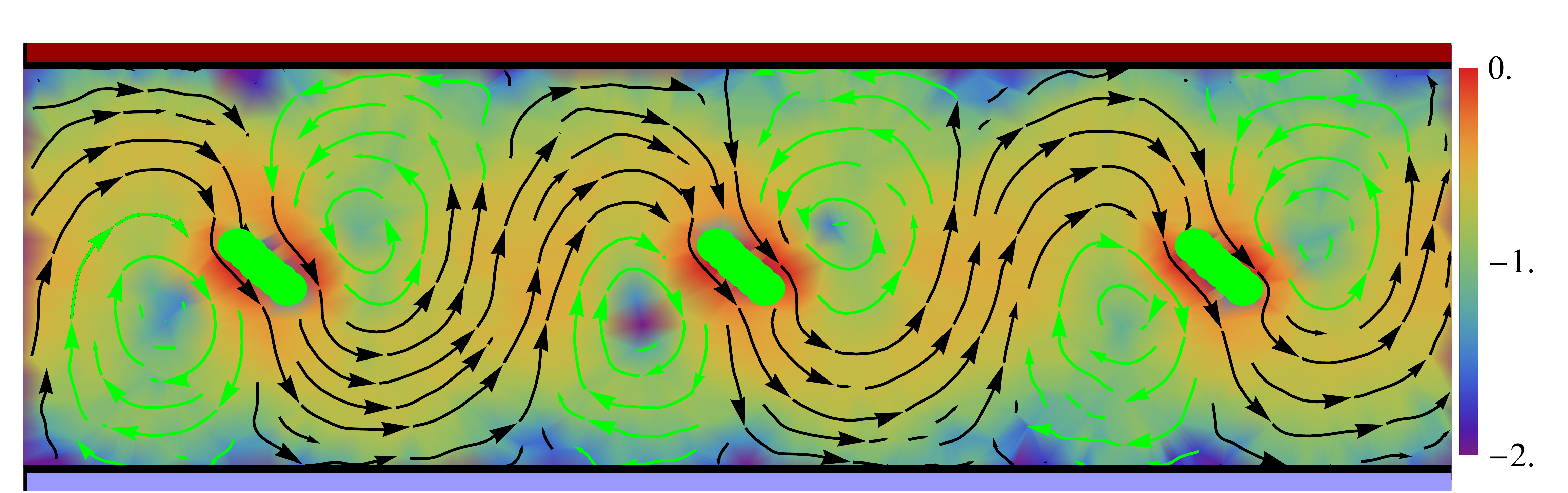}
\caption{Flow streamlines corresponding to the micropump in
  Fig.~\ref{schemes}(b). Here and in what follows, the background
    color codes the magnitude of local velocities $|\textbf{v}(x,z)|$
    rescaled with a factor of
    $\log(|\textbf{v}|-|\textbf{v}|_{min})/\log(|\textbf{v}|_{max}-|\textbf{v}|_{min})$,
    where $|\textbf{v}|_{max}$ and $|\textbf{v}|_{min}$ are maximum
    and minimum of magnitudes of the velocities in each vector
    field.  As a guide to the eye, black lines correspond to the
      net flow, and green lines to the vortexes.}
	\label{fig:sch_flow}
\end{figure*}

A representative simulation output of the temporally averaged flow
streamlines of a cross-section in $x-z$ plane with three pillars is
illustrated in Fig.~\ref{fig:sch_flow}. In this flow pattern an
  ``effective flow'' region where the fluid passes close to the
  obstacles and continues along the microchannel with a
  sinusoidal-like trajectory. Additionally ``vortex'' region can be
  identified, where the flow rotates without providing any net
  contribution to the total flux. %%%

\begin{figure}[h]
  \topinset{ (a) }{
	\includegraphics[trim={0 0 -0.5cm 0},clip,scale=0.11]{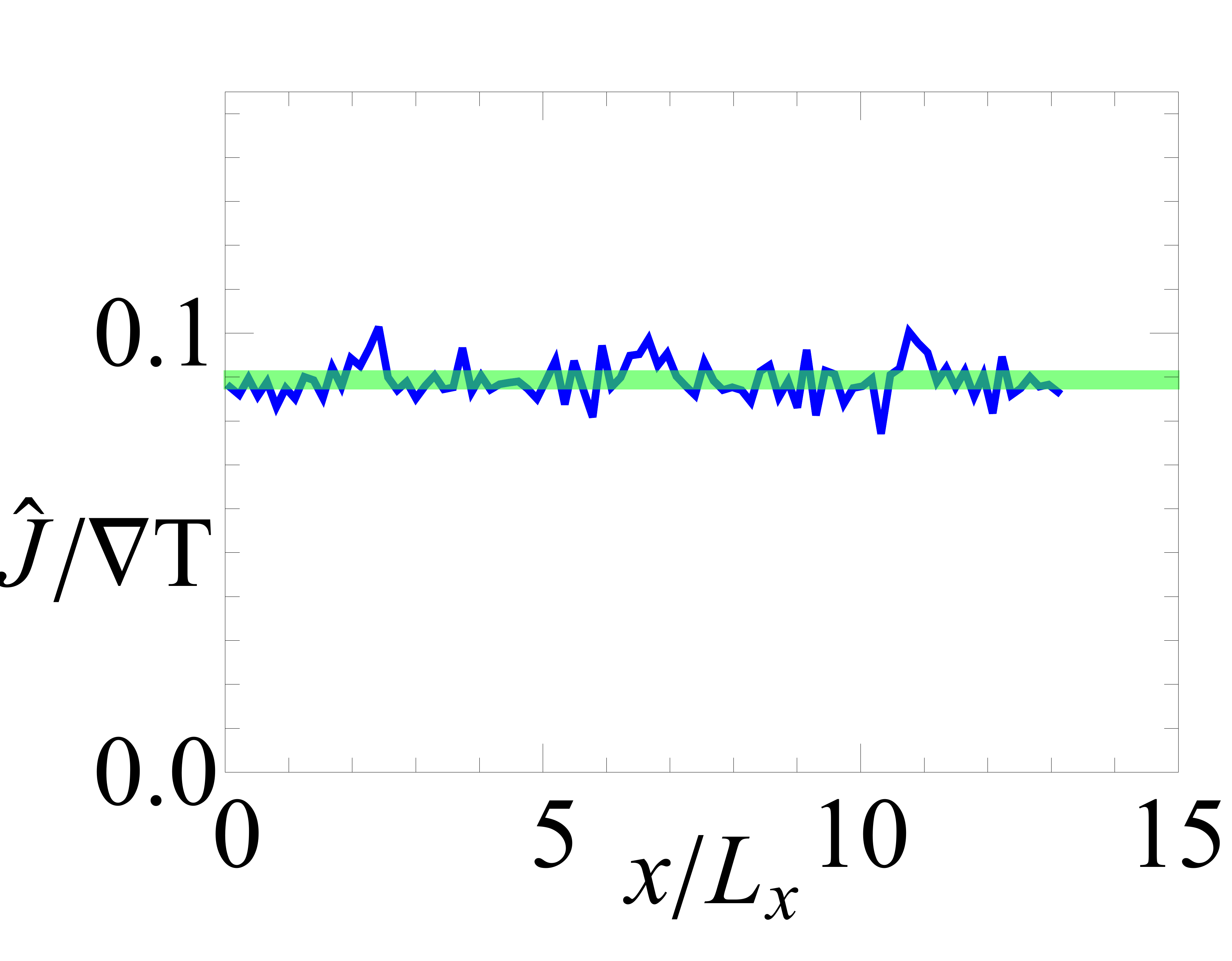}}{0.14in}{-0.35in}
  \topinset{ (b) }{
	\includegraphics[trim={0 -0.5cm 0 0},clip,scale=0.117]{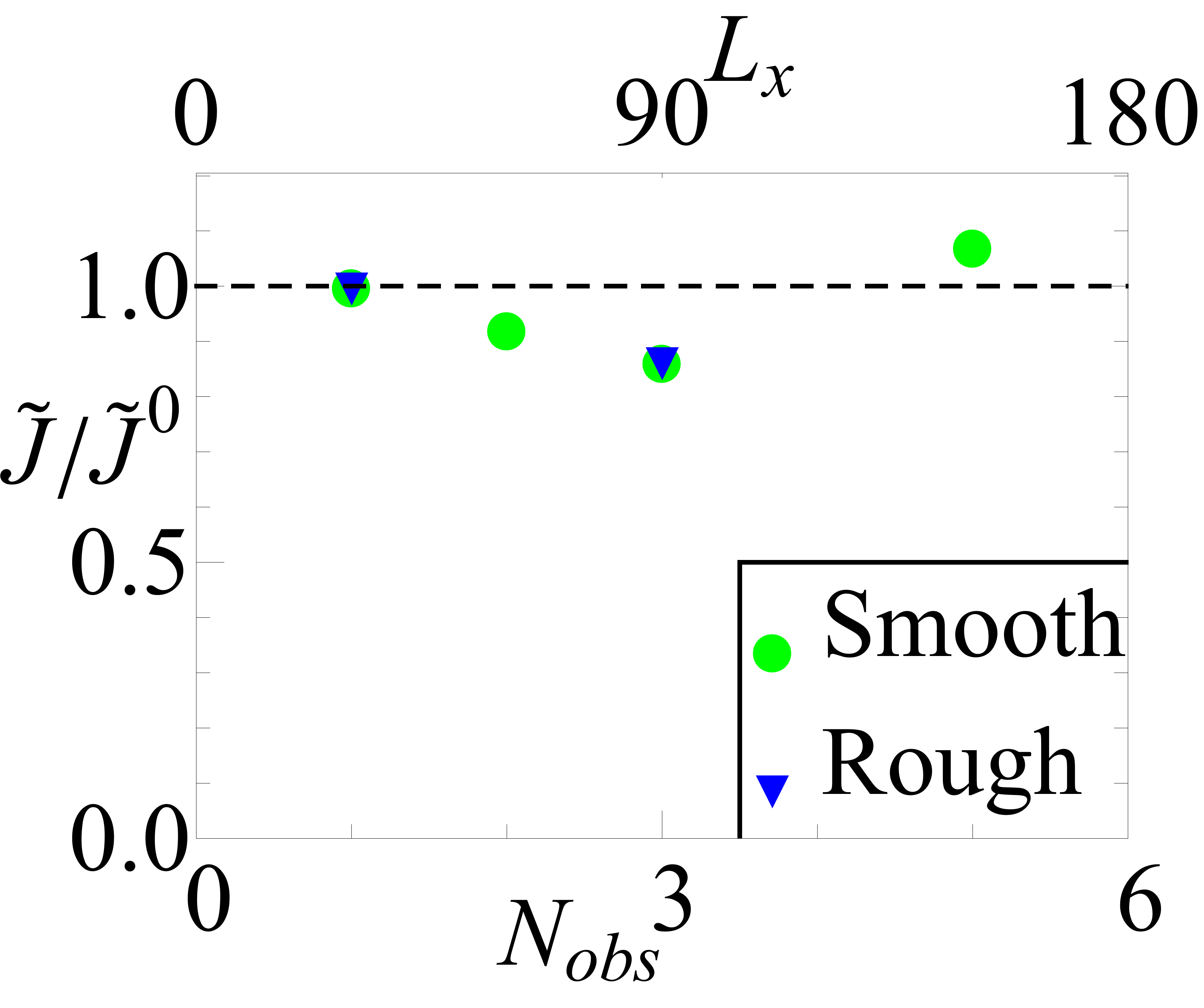}}{0.22in}{-0.35in}
	\caption{(a)~Scaled flux density corresponding to the setup in
          Fig.~\ref{fig:sch_flow}, calculated along the channel
          length. The green transparent line represents the average value. (b)~Normalized flux density calculated with
          different number of pillars in the primary simulation box,
          $N_{obs}$, with constant obstacle separation
          $D$. $\tilde{J}^0$ is the flux density with one obstacle in
          the primary box, this is with $D=L_x$, for two different
          values of the pillar rugosity.} 
	\label{Flux}
\end{figure}

The resulting net flux density $\hat{J}$ as defined in
Eq.~(\ref{flux}) is computed along the channel, and displayed in
Fig.~\ref{Flux}(a) normalized by the externally applied temperature
gradient. In spite of the stream lines tortuosity, and due to the mass
continuity, the flux density is basically constant along the channel
(see Fig.~\ref{Flux}(a)). Note that the flow does not depend on the
position in the $y$ axis, such that we average the flow field in this
direction to increase the statistics accuracy. Both the flow field and
the flux at steady state are temporal average over $5\times 10^5$
units of time with at least $24$ simulation measurements. Note that
increasing the number of simulations would eventually improve the
statistical dispersion of the data, and decrease the (not indicated)
error bars, but it would not change any of the presented conclusions.

Given the system symmetry and the use of PBC along the channel, the
fluid flow is not expected to significantly depend of the number of
pillars considered, for a fixed interparticle separation $D$.
However, PBC might also lead to fluid correlations, which would be
enhanced for smaller system sizes~\cite{huang12}.  In Fig.~\ref{Flux}(b)
measurements of the flux densities are shown for simulations with
different number of obstacles in the primary simulation box,
$N_{obs}$, where the box dimension has been accordingly varied
$L_x=N_{obs}D$.  The flux densities show to have a relatively constant
value, independent on the system size, and furthermore flow patterns
around each pillar are identical for different $N_{obs}$ which
allows us to investigate systems with just one pillar without
diminishing the applicability of the conclusions.

\subsection{Interfacial properties}

\begin{figure}[h]
\centering
\includegraphics[height=0.27\textwidth]{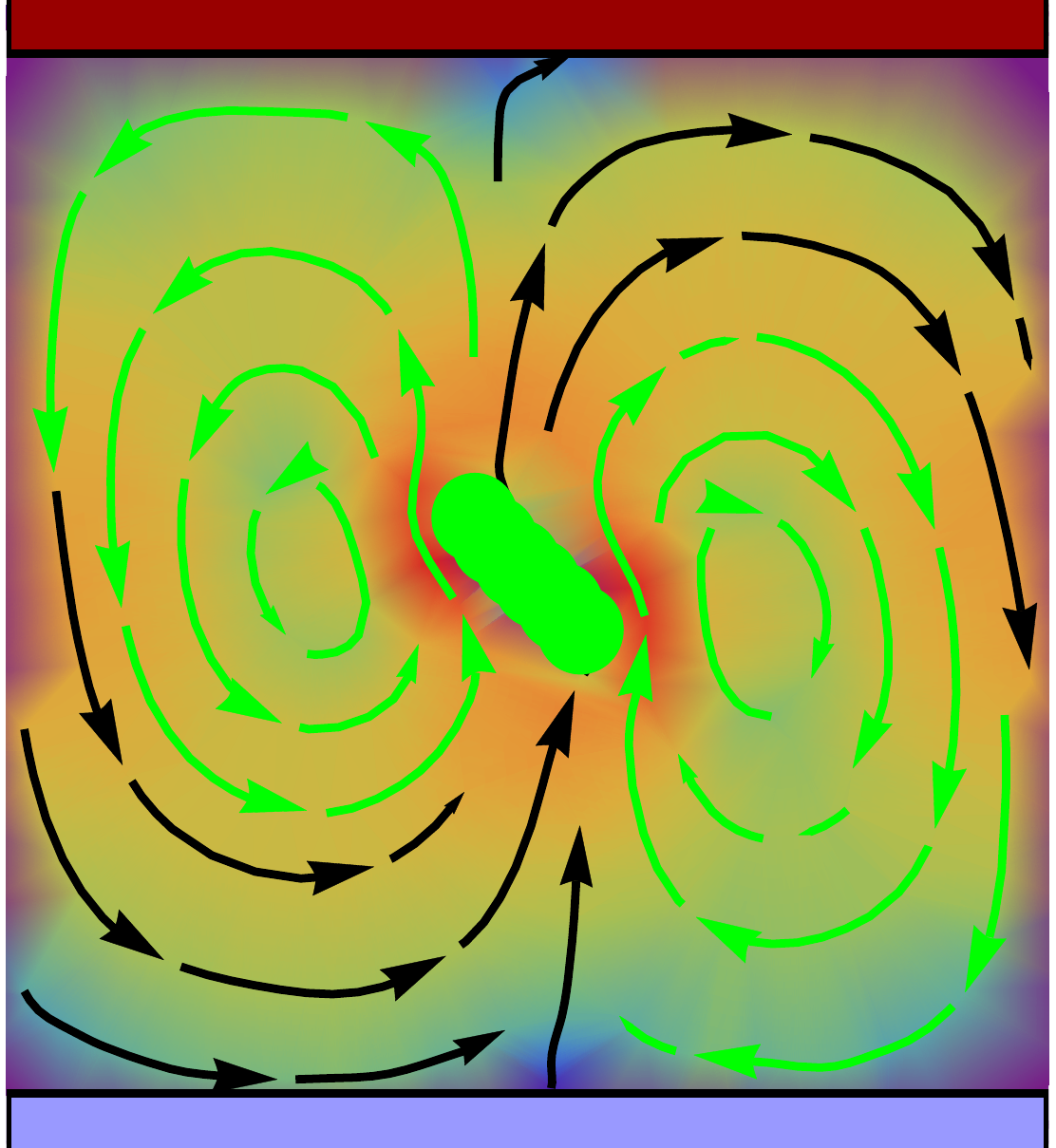}
\caption{Flow streamlines of a micropump with a thermophobic pillar
  ($a24$ potential).}
	\label{fig:rep_att}
\end{figure}

The thermophilic character of the pillar in Fig.~\ref{fig:sch_flow}
can be observed in the flow close to the obstacle surface, where the
flow is clearly directed from warm to cold areas, opposite to the
thermophoretic force on the pillar beads, as
expected~\cite{yang13flow}. A similar micropump, with an obstacle
constructed out of thermophobic beads is displayed in
Fig.~\ref{fig:rep_att}, where the flow directed then to the warm areas
can be observed in the neighbourhood of the obstacle surface. In spite
of this difference, the overall direction of the flux is the same in
these two cases, which results in flow streamlines with significantly
different pathways in both cases, as can be seen by the flow close to
the pillar and the position of the stagnation point relative to the
obstacle.

\begin{figure}[h]
\centering
	\includegraphics[width=0.4\textwidth]{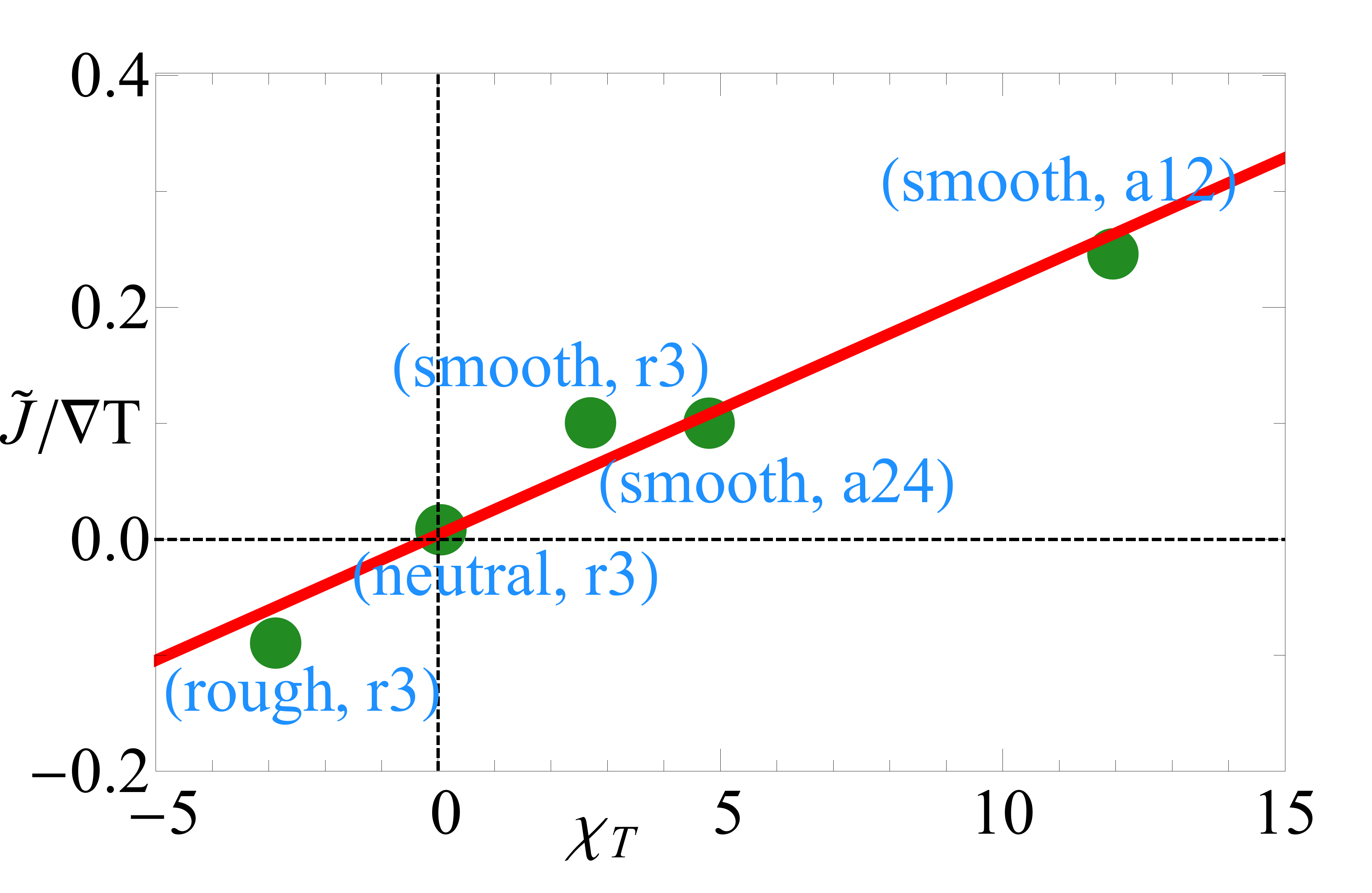}
	\caption{Variation of the normalized flux
          $\widetilde{J}/\nabla T$ as a function of the anisotropic
          thermophoresis factor $\chi_T$, with parameters specified in
          table~\ref{tb:chit}. Red line correspond to linear behavior
          in Eq.~(\ref{eq-jxa}).}
	\label{fig:chit}
\end{figure}

As stated in Eq.~(\ref{eq-jxa}) the overall direction of the flow
parallel to the walls is determined by the anisotropic factor $\chi_T$
in Eq.~(\ref{chi}). This explains that pillars with thermophilic and
thermophobic character might still result in flux with the same
direction. To further verify this statement we have quantified
$\chi_T$ in some cases by additional simulations with obstacles placed
parallel and perpendicular to the channel walls, with a procedure
similar to the one described in Ref.~\cite{ZT01}. Results and
simulation parameters are summarized in Table~\ref{tb:chit}, and the
variation of the flux normalized by the applied temperature gradient
with the corresponding anisotropic thermophoresis factor $\chi_T$ is
displayed in Fig.~\ref{fig:chit}, where the linear behavior predicted
by Eq.~(\ref{eq-jxa}) is nicely confirmed. Note that in our
simulations the pillars along $y-$direction are constructed by
PBC. Therefore the values of $\chi_T$ are represented as the measured
$\chi_T$ per unit length along $y-$axis.

\begin{table}[h]
	\begin{tabular}{|l || c | c | c | c | c|}
		\hline  
		potential & $r3$  & $r3$ & $r3$ & $a24$ & $a12$ \\
	$l/d$ & $1.0$  & $0.8$ & $0.4$  & $0.4$ & $0.4$ \\
	$W/d$ & $3.60$  & $3.08$ & $2.93$  & $2.93$ & $2.93$ \\               
		$\nabla T$ & $0.0207$ & $0.0138$ & $0.0138$ &  $0.0172$ &  $0.0172$ \\
		\hline  
		$\alpha_{T,\perp}$ &  $-9$ & $-10$ & $-13$ & $27$  & $44$ \\
		$\chi_T$ &  $-2.88$ & $0.050$ & $2.700$ & $4.800$  & $11.950$ \\
		$\tilde{J}/\nabla T$ & $-0.086$ & $0.011$ & $0.104$ &  $0.103$&  $0.249$\\
                \hline
	\end{tabular}
	\caption{Measurements of $\chi_T$ and flow fluxes $\tilde{J}$ for the simulations performed with 
          the default geometrical dimensions, namely $H=D=L_x=12d$.}
	\label{tb:chit}
\end{table}

\begin{figure}[!ht]
	\hspace{-1.2cm}
%	\centering
	\mbox{\subfloat{	
		\begin{picture}(150,150)
		\put(0,0){\includegraphics[height=0.27\textwidth]{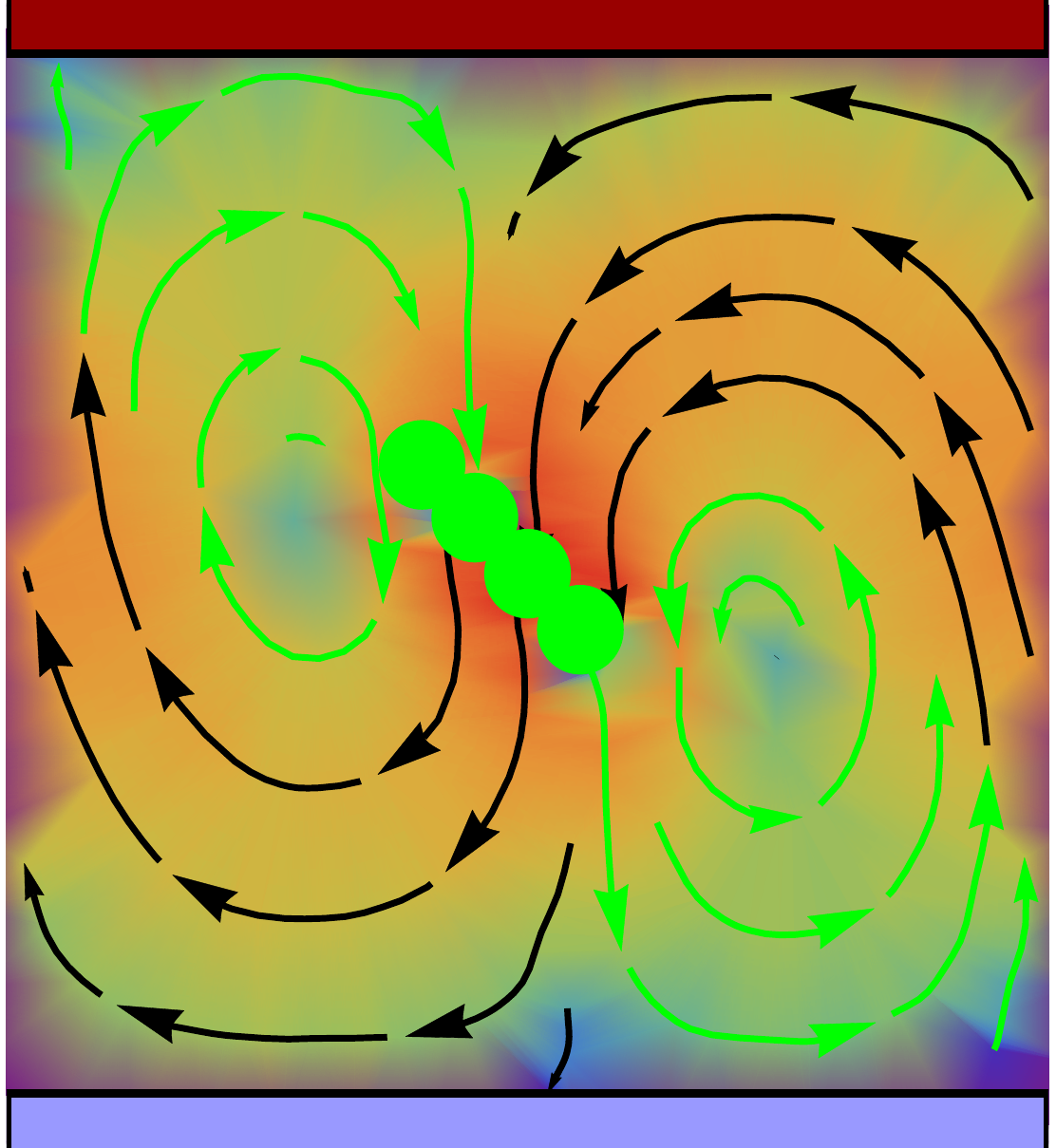}}
		%\put(55.0,110){\textcolor{white}{\Large{rough}}}
		\put(12.0,120){\textcolor{white}{(a)}}
		\end{picture}	
		\label{rough}
	}
	\hspace{-0.8cm}
	\subfloat{		
		\centering
		\begin{picture}(150,150)
		\put(0,0){\includegraphics[height=0.27\textwidth]{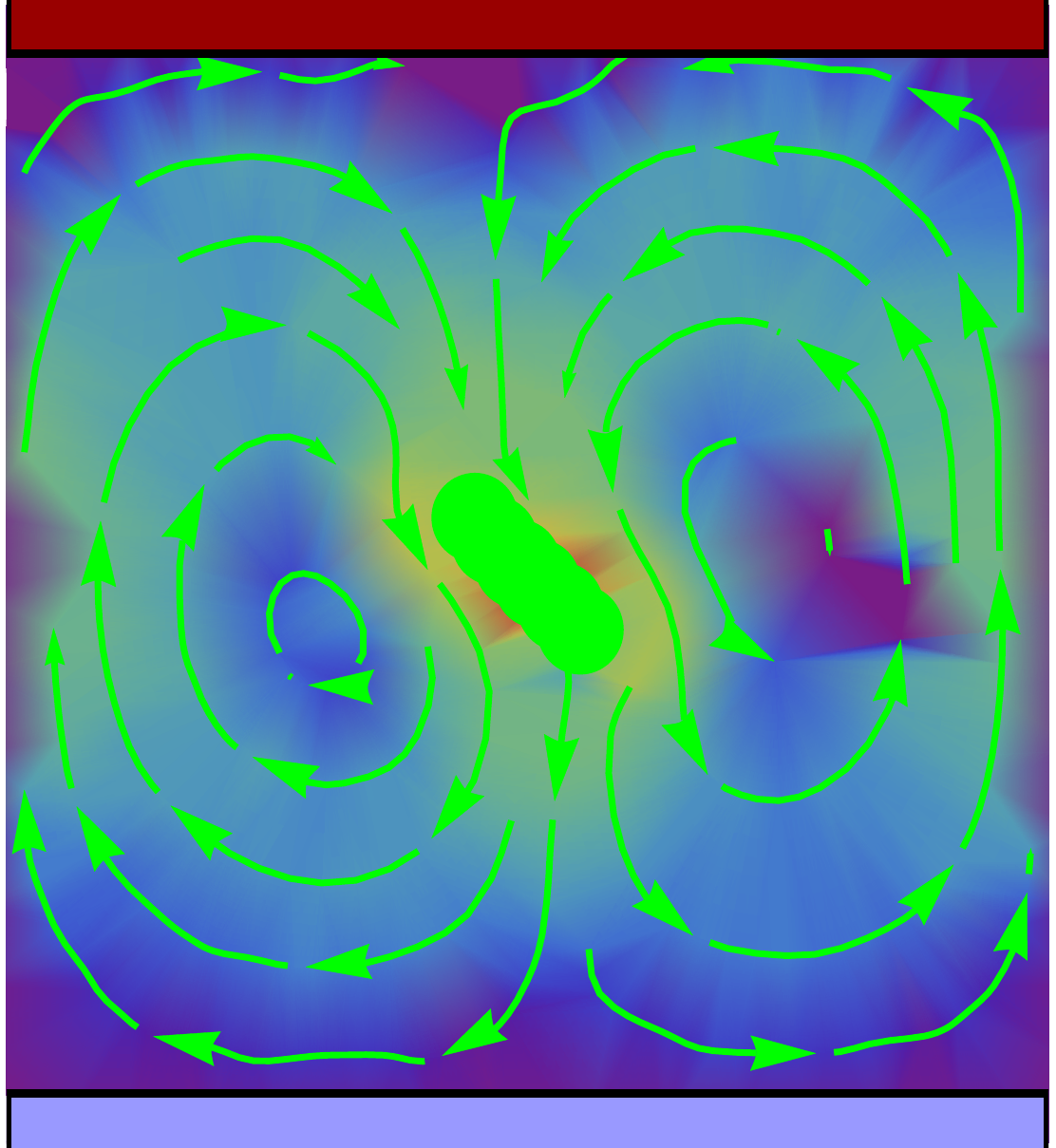}}
		%\put(55.0,70){\textcolor{white}{\Large{neutral}}}
		\put(12.0,120){\textcolor{white}{(b)}}
		\end{picture}
		\label{neutral}
	}}
      \caption{Flow fields with varying pillar surface rugosity:
        (a)~rough obstacle surface, $l/d=1.0$. (b)~neutral obstacle
        surface, $l/d=0.8$.}
	\label{figrugosity}
\end{figure}

The investigation of the anisotropic thermophoretic effect in
colloidal rods~\cite{ZT01} showed a sign change of $\chi_T$ with the
rod rugosity.  We also explore the dependence with rugosity in the
case of elongated obstacles as depicted in
Fig.~\ref{schemes}\subref{fig:obstacle}, and the explicit measurements
of $\chi_T$ shown in Table~\ref{tb:chit} display similar sign
change. The streamlines for microchannels with elongated rough and
neutral thermophilic pillars are plotted in Fig.~\ref{figrugosity},
while the smooth corresponding case can be seen in
Fig.~\ref{fig:sch_flow}.  The flux direction, as well as the flow
pathways are different in the three cases as a consequence of the
different values of $\chi_T$. 
%cross through the interstices created by the large separation between
%beads, which is not the only reason for the negative value of
%$\chi_T$, since for example in the case of the thermophobic obstacle
%in Fig.~\ref{fig:rep_att} $|\alpha_{T,\|}|<|\alpha_{T,\perp}|$, which
%might be considered less intuitive. Practically, this case can be
%  viewed the pillar is made of loose or porous material and fluid can
%  still pass through.%%
Besides the change in  flux direction, the location of the
vortexes and the corresponding stagnation points vary also with
$\chi_T$ and the thermophoretic character.  For the microchannel with
smooth thermophilic obstacles (Fig.~\ref{fig:sch_flow}), this is when
$|\alpha_{T,\|}|>|\alpha_{T,\perp}|$, the stagnation points are
aligned perpendicular to the long pillar axis. On the other hand,
for cases with $|\alpha_{T,\|}|<|\alpha_{T,\perp}|$, these are smooth
thermophobic (Fig.~\ref{fig:rep_att}) or rough thermophilic obstacles
(Fig.~\ref{figrugosity}(a)); the stagnation points are aligned close to
the long pillar axis. %%
Finally, it is interesting to note that there is an intermediate
rugosity case, which we call neutral, for which the anisotropic effect
vanishes, $\chi_T \simeq 0$. This is the case shown in
Fig.~\ref{figrugosity}(b) where the flow field shows a symmetric pattern
with very close to vanishing flux, which resembles the one induced by
a fixed isotropic colloid~\cite{yang13flow}.

\subsection{Channel geometrical properties}

Besides the interfacial properties discussed until now, the channel
dimensions are naturally going to affect the flow intensity, shape,
and exceptionally even its sign, as indicated by the prefactor in
Eq.~\eqref{eq-jxa} being function of the channel width, pillar
interseparation, and obstacle aspect ratio.

\paragraph{Channel width effect}
%\paragraph{Transverse direction: effect of channel width}

\begin{figure}[h]
	\centering
			\subfloat{
				\centering
				\def\stackalignment{l}
				\topinset{ \textcolor{white}{(a)} }{\includegraphics[height=0.165\textwidth]{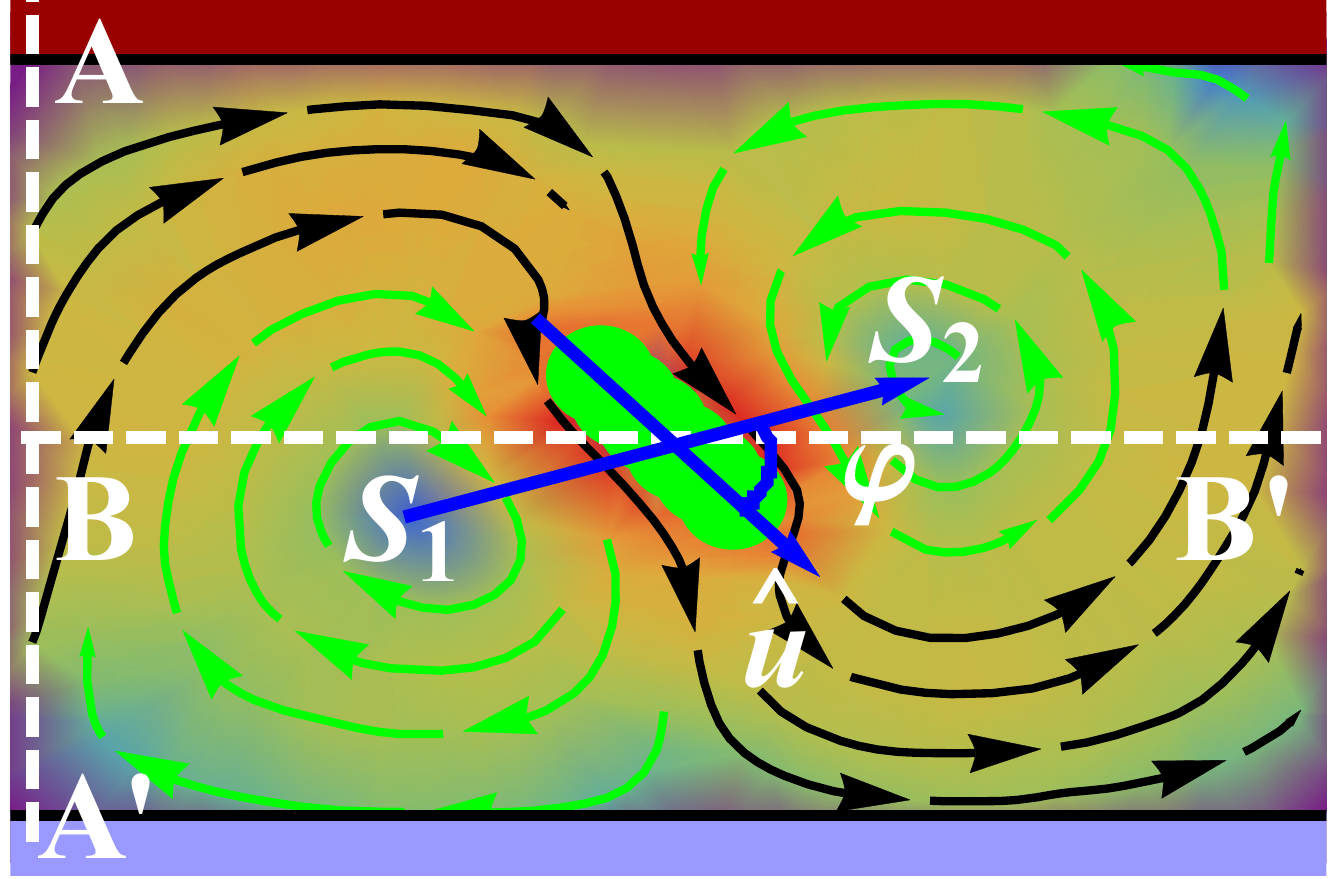}}{0.2in}{0.05in}  
				%	\caption{}	\includegraphics[width=6cm,height=3.5cm]{H-20.pdf}
				\label{fig:H20}     }\\
			\subfloat{	
				\def\stackalignment{l}
				\topinset{ \textcolor{white}{(b)} }{\includegraphics[height=0.5\textwidth]{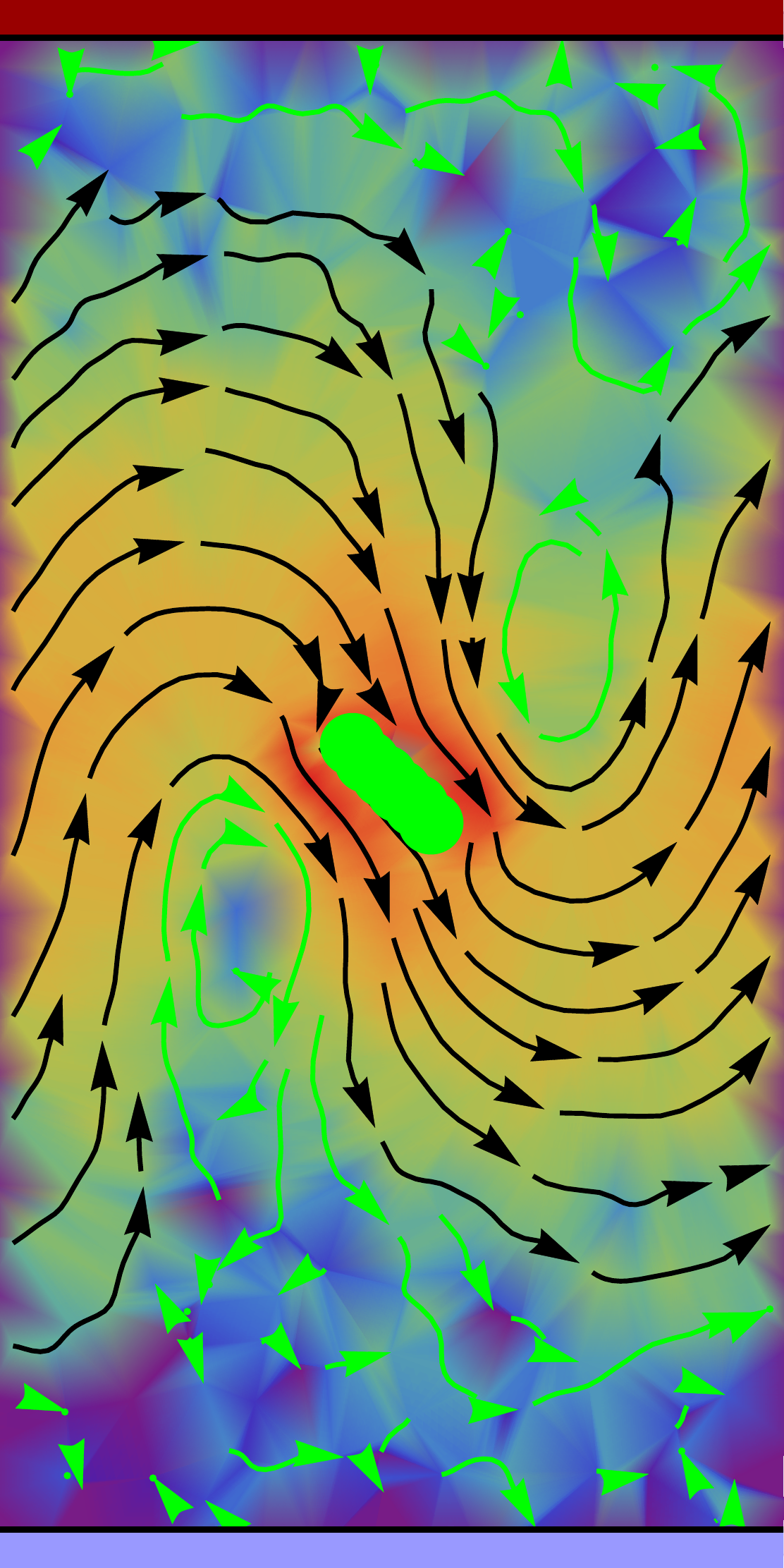}}{0.2in}{0.05in}   
				%	\caption{}	\includegraphics[width=6cm,height=9cm]{H-60.pdf}
				\label{fig:H60}  }			
	\caption{Flow field for channels of different channel widths. (a)~Strongly confined case with $H=2.93W$. 
                 This figure shows the description of the stagnation points with the associated angle 
                 $\varphi$ and $S_d=\left|  \protect  \overrightarrow{S_1S_2}  \right|$ the stagnation separation. (b) Loosely confined case with $H=8.78W$.}
	\label{Lz}
\end{figure}

\begin{figure}[h]
\hspace*{-0.5cm}
\includegraphics[width=0.5\textwidth]{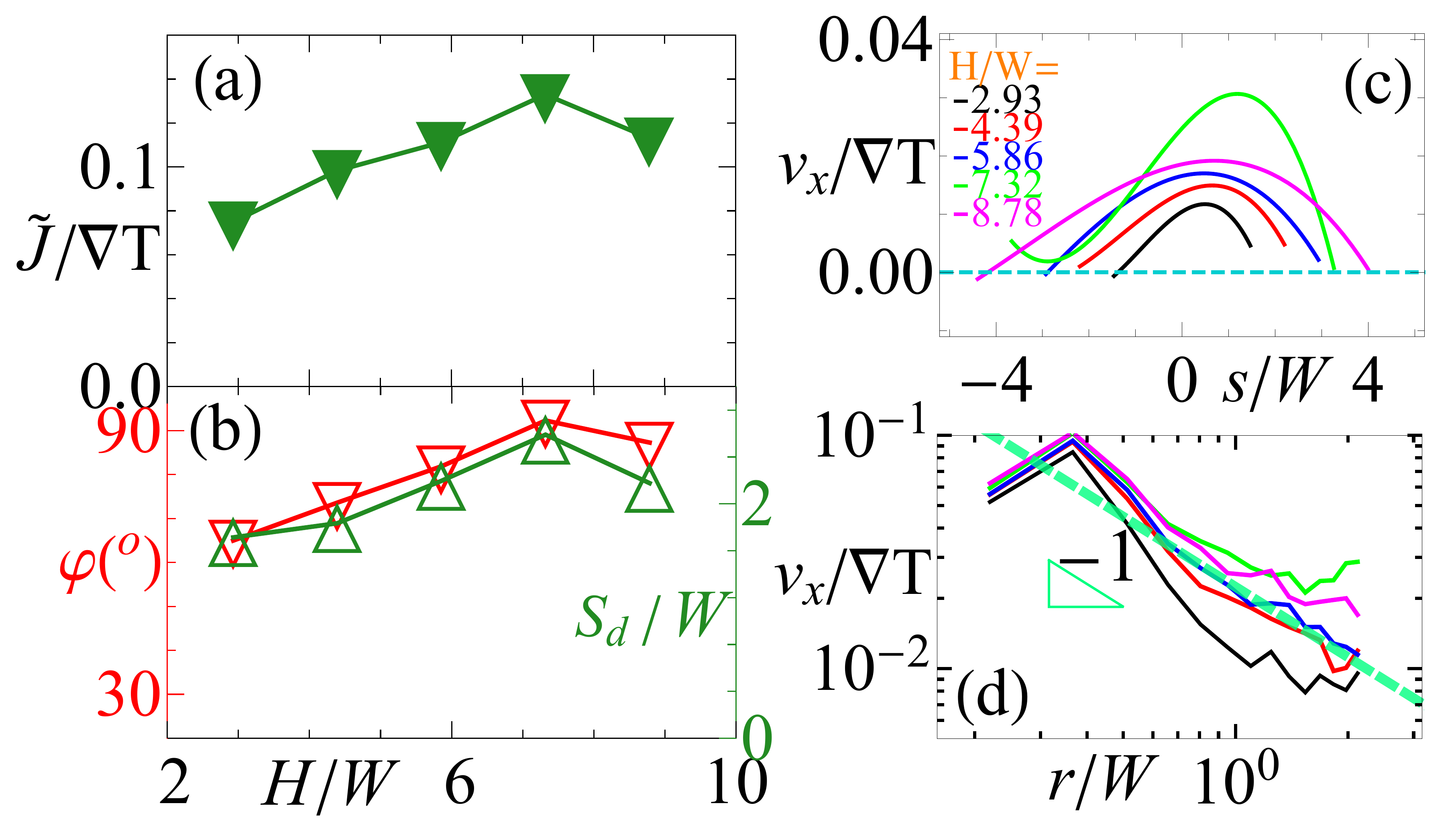}
%\raggedright
\caption{(a) $\widetilde{J}/\nabla T$ as a function of channel width,
  $H$. (b) Stagnation angle $\varphi$, and distance $S_d/W$ as a
  function of $H$. (c) Velocity profiles along the channel direction
  $v_x$ in the axis perpendicular to the channel and exactly in
  between two pillars ($A-A'$ in Fig.~\ref{Lz}(a)), for various
  channel widths, shown as a polynomial fit to the data. $s$ is the
  distance to the middle of the channel. (d) $v_x$ profile in an axis
  parallel to the walls crossing the middle of the pillar ($B-B'$ in
  Fig.~\ref{Lz}(a)), as a function of $r$, the separation to the
  obstacle center.}
	\label{vx}
\end{figure}

To study the effect of the channel width on the induced flow, we
perform simulations varying $H$, approximately from $3$ to $9$ 
times the obstacle length $W$, and fixing all the other dimensions to
default values, as well as the temperatures at the two walls
($T_h=1.31$ and $T_c=0.68$).  Figure~\ref{Lz} shows the cases with
largest and smallest confinements here investigated, which can be
compared with the intermediate case in Fig.~\ref{fig:sch_flow}. %
The confinement provided by the no-slip channel walls restricts the
effective flow domain and adjusts the fluid flow perpendicular to the
walls rather than parallel to them.
This wall-restriction of the flow explains the linear increase of
the flux with channel width shown in Fig.~\ref{vx}(a). For channels wide
enough, the flow close to the walls is weak, close to
$10^{-2}\left|v\right|_{max}$, such that confinement does not
noticeably affect anymore the flow, and an increase of the channel`
width results in a straightforward decrease of the net flux. This
explains the maximum observed in Fig.~\ref{vx}(a) for channels with $H
\simeq 7.3W$.
%%%

Additionally, Fig.~\ref{Lz} shows that the channel width also
importantly changes size, shape, and location of the vortex regions,
which is a consequence of the previously discussed flow distortion.
The vortexes can be quantitatively characterized with the location of
the middle points of zero flow velocity, by defining the stagnation
angle $\varphi$ and the normalized stagnation distance $S_d/W$. As
indicated in Fig.~\ref{Lz}(a), $S_d$ is the distance between the two
stagnation points $S_1$ and $S_2$ around a given pillar, and $\varphi$ is the angle
that the line connecting these two points makes with the pillar
elongated axis. Fig.~\ref{vx}(a) shows that both these quantities have a
maximum at the same point as the normalized flux, and that this occurs
when $\varphi=90^\circ$, this is when the vortex middle points are
exactly perpendicular with the elongated obstacle.  When $H$
is as small as shown in Fig.~\ref{Lz}(a), the vortex region takes up
large part of the microchannel due to the significant confinement in
the gradient direction, and the distance between complementary
stagnation points is very small. The size and position of the vortex
changes exactly until $\varphi=90^\circ$, value from which the
vortexes only distort due to the lack of confinement. 

The velocity profiles along the channel direction $v_x$ in the
cross-sections perpendicular and parallel to the channel walls are
shown in Fig.~\ref{vx}(c) and (d), respectively.  Figure~\ref{vx}(c)
shows that the velocity profile in between the obstacles is close to
parabolic, with a slightly off-center maximum, in which the maximum
values of the velocity follows the same dependence as the fluxes in
Fig.~\ref{vx}(a).  This means that the intensity of the flow
  decays close to the walls, and this decays is then larger for the
  widest channels where there can even be a residual flow in opposite
  direction.  The velocity profile along the middle axis parallel to
the walls is shown in Fig.~\ref{vx}(d).  The flow field in the
most immediate proximity of the pillar could be understood as
determined by the intrinsic obstacle properties and only weakly
modified by the boundary conditions. The velocity profiles in
Fig.~\ref{vx}(d) show a $1/r$ decay as the distance from the pillar
surface increases. This is similar to the flow for a fixed phoretic
colloid, and has the same origin. Deviations from this behavior are
due to the confinement.

We have analyzed the dependence of the channel width for just
one parameters set, and fixed interfacial properties; we expect almost
identical behaviors for other relevant parameter sets, although the
exact location of the optimal channel width, here $H/W\approx7.3$
could eventually vary.

\begin{figure}
  \topinset{  \textcolor{white}{(a)}}{\includegraphics[height=0.27\textwidth]{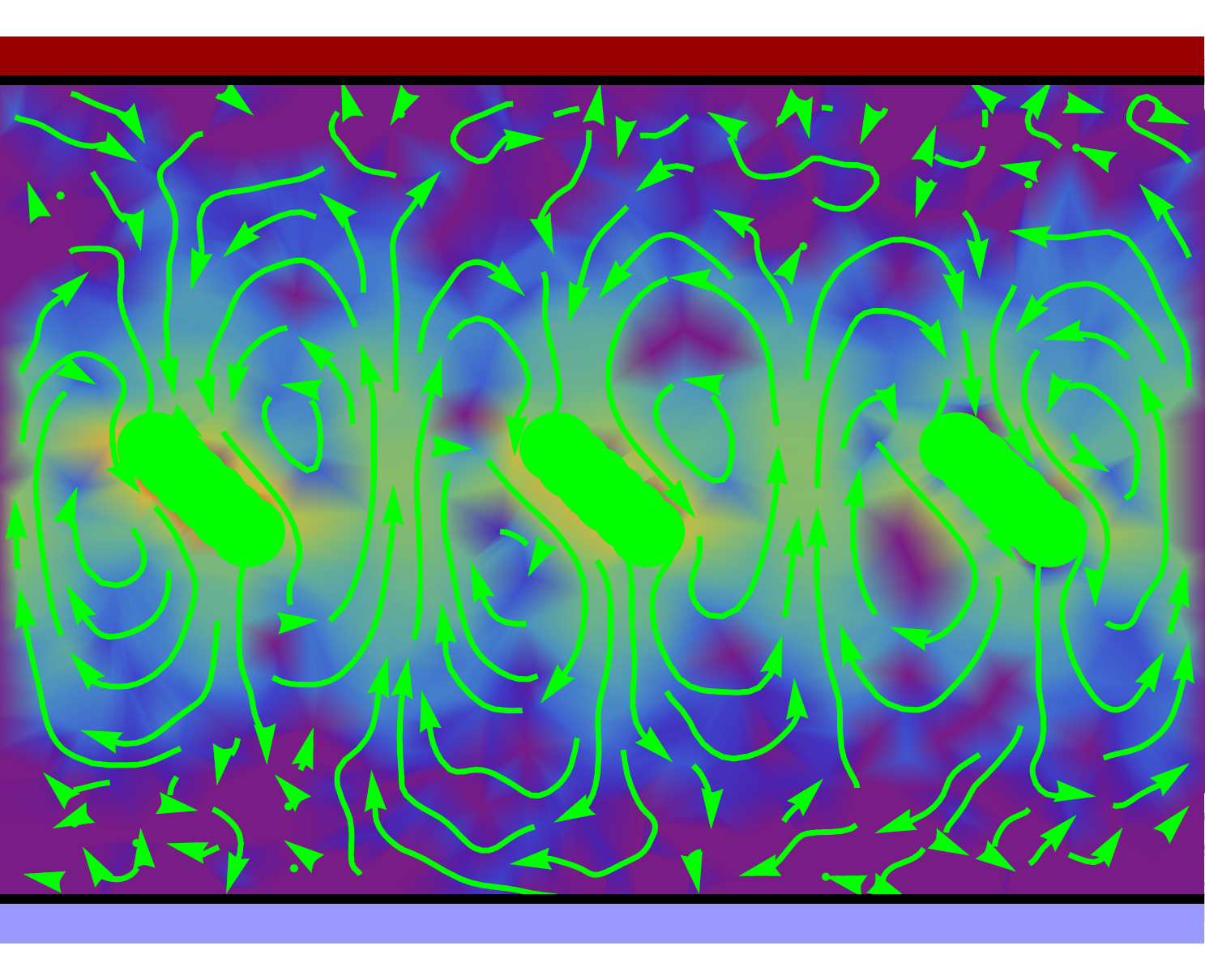}}{0.2in}{-1.1in}
  \topinset{ \textcolor{white}{(b)} }{\includegraphics[height=0.27\textwidth]{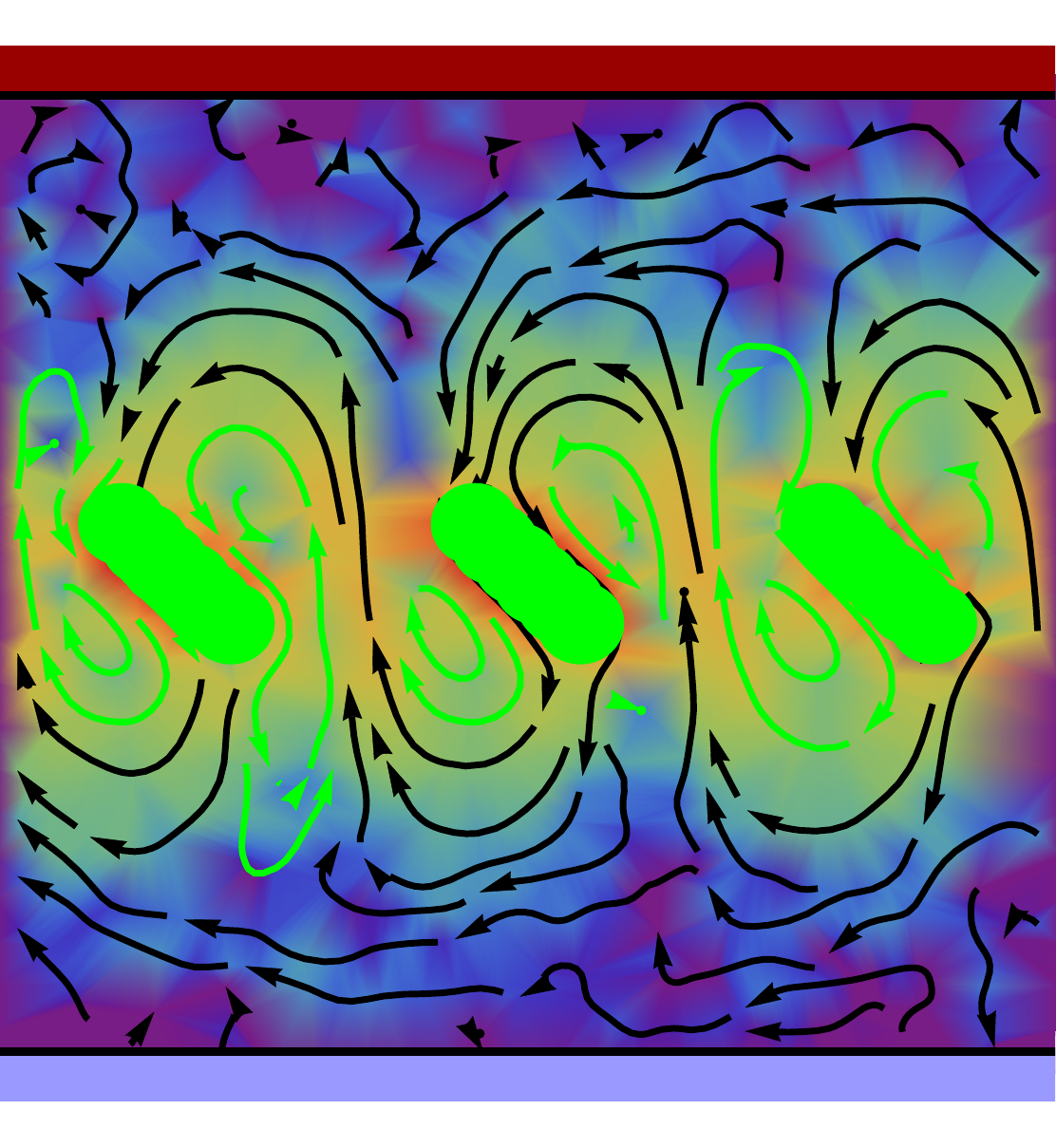}}{0.2in}{-0.8in}
  \caption{Flow patterns variation with very small obstacles separation
    $D$. (a)~Vanishing flux case for $D=1.95W$, (b)~Reversed flux case for $D=1.46W$.}
	\label{fig:Dflow}
\end{figure}

\paragraph{Inter-obstacle  separation effect}
In the limit of vanishing separation between obstacles, there is no
expected net flux generated by the micropump; and in the complementary
limit, one single pillar will not be able to create a considerable
flux in an arbitrary long channel. To most precisely understand the
transition between these limits, simulations at various pillar
separations have been carried out. %%
Figure~\ref{fig:Dflow} shows that the relevance of the vortex area
and the tortuosity of the streamlines decreases with increasing
obstacle separation. This can be understood since the flow field in
the most immediate proximity of the obstacle has an intensity and
direction determined by the surface rod properties, while the flow
slightly further away from the rod, just needs to adapt to the given
boundary conditions.

\begin{figure}
\hspace*{-0.5cm}
%\raggedright
 \includegraphics[width=0.5\textwidth]{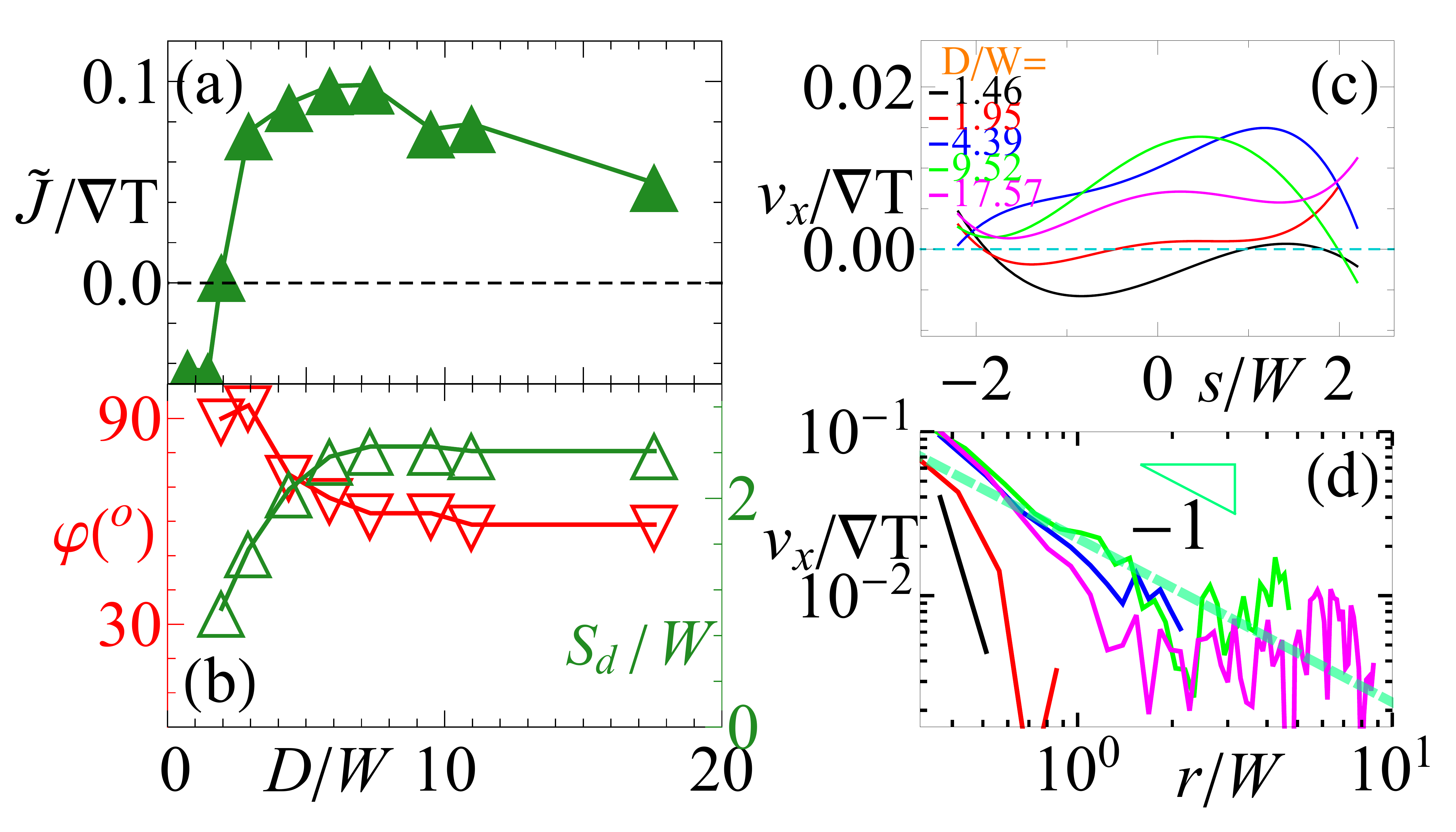}
  \caption{(a)~$\widetilde{J}/\nabla T$ dependence with
    obstacle inter-separation distance $D$.
    (b)~Corresponding $\varphi$ and $S_d$ as a function of
    $D$. (c) and (d) Velocity profiles, similar to
    Fig.~\ref{vx}(c) and (d).}
  \label{fig:Dflux}
\end{figure} 

% This occurs when the relative size of the vortex area is larger
% than the previously called effective area, and it is due to the
% fact that the direction in both areas has opposite directions.
Decreasing the pillar separation importantly distorts the vortex
areas, such that the counter-rotating flow becomes relatively more
important.  This makes that the net flux importantly decreases for
small (but not vanishing) pillar separation reaching a vanishing value
as shown in Fig.~\ref{fig:Dflow}(a), or even negative as shown in
Fig.~\ref{fig:Dflow}(b) and Fig.~\ref{fig:Dflux}(a). %%
Increasing obstacle separation increases the flux, see
Fig.~\ref{fig:sch_flow}, until a maximum value is obtained ($D/W
\approx 7$, for the parameters here employed), as shown in
Fig.~\ref{fig:Dflux}(a). Increasing the separation further than
  this optimal value makes the flux decrease, but interestingly, this
decay occurs very slowly. In this way, for the case with the largest
separation where simulation has been here performed ($D/W\approx18$)
the total flux is just $50\%$ smaller than that with the optimal
separation. This result can be practically advantageous since devices
with a smaller number of pillars will typically be easier and
therefore cheaper to produce.

The location of the vortexes center, which can be inferred from
$\varphi$ and $S_d$ in Fig.~\ref{fig:Dflux}(b), changes with
increasing obstacle separation until they reach a stable location.
This location does not depend anymore on $D$, and seems therefore to
be determined by the value of the channel width. The velocity profiles
in the cross-section in between pillars is shown in
Fig.~\ref{fig:Dflux}(c) for various $D$ values. The profiles are
naturally related to their fluxes, such that small separations show
flat profiles which can be averagely negative, vanishing or positive,
while larger separations show progressively more parabolic-like
profiles slightly tilted towards the pillar direction. %%
Note that the flow induced by this micropump is intrinsically
  non-parabolic, see for example Fig.~\ref{fig:sch_flow}. The observed
  close-to-parabolic profiles occur just in the perpendicular axis, in
  between the pillars for intermediate separations.

The flow parallel to the walls at the center of the channel is shown
in Fig.~\ref{fig:Dflux}(d), where the flow field in the most immediate
proximity of the obstacle shows a bit stronger dependence with $D$
than with $H$ in Fig.~\ref{vx}(d), but with similar $1/r$ decay. Note
that the velocity profiles for the smallest inter obstacle separation
deviate from the $1/r$ decay scenario since their effective flow
patterns are too strongly distorted.

\begin{figure}
  \includegraphics[height=0.27\textwidth]{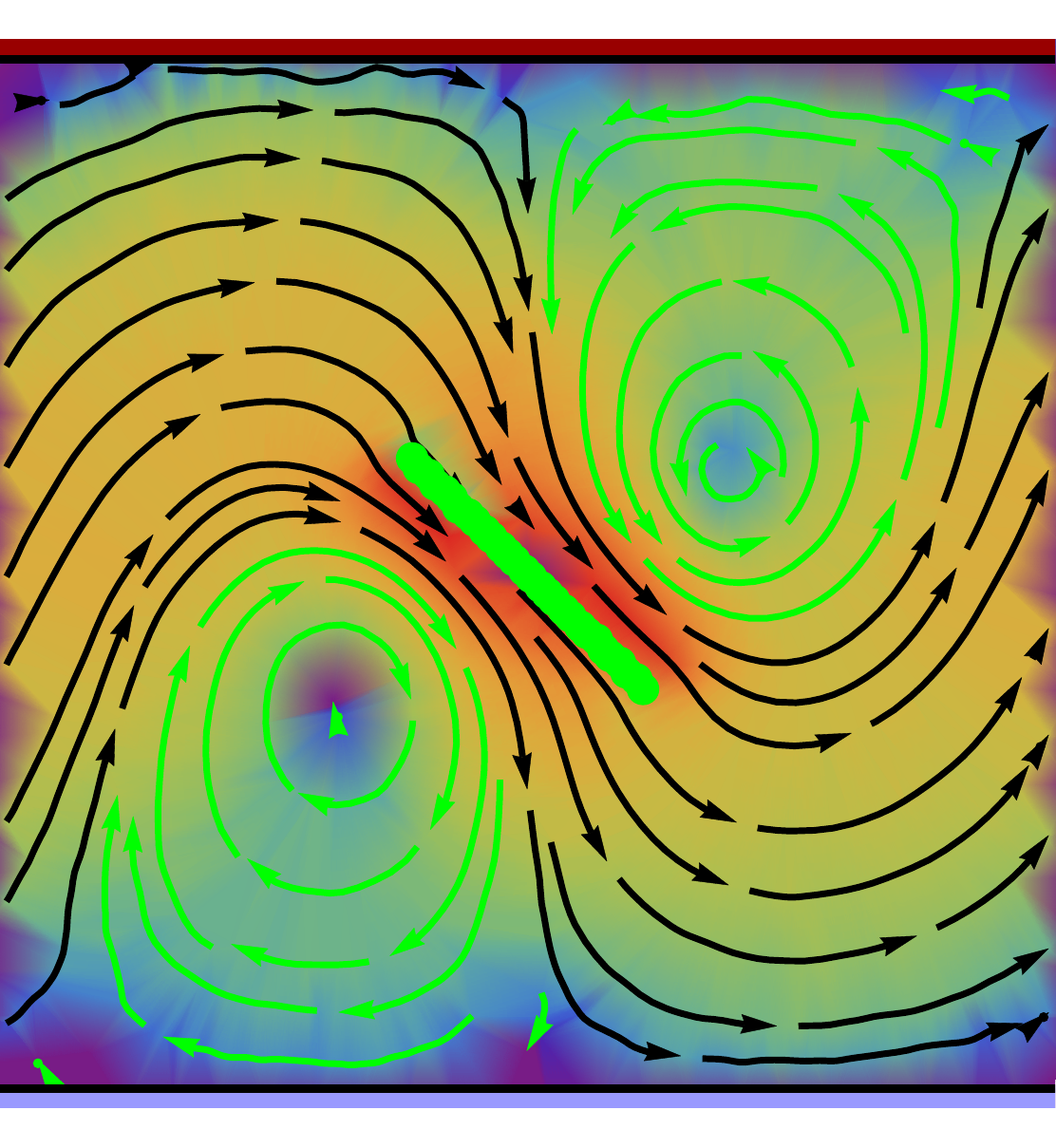}
  \caption{Flow streamlines in a microchannels with an obstacle with
    large aspect ratio, $W/d \simeq 11$. }
  \label{fig:ASP}
\end{figure}

\paragraph{The role of the obstacle aspect ratio}
As already described for the case of rodlike colloids~\cite{ZT01}, the
anisotropic thermophoretic factor of elongated obstacle is expected to
increase linearly with aspect ratio, and therefore the driving force
along the channel for the anisotropic thermophoretic pump here
investigated. In order to investigate the effect of the aspect ratio,
and reasonably decouple it from the other geometrical effects, we
perform simulations varying channel width and inter-pillar
separation by keeping fixed ratios, $D/W=2.93$ and $H=D$, and although
its effect can be understood with proper normalization, we here vary
the wall temperatures to fix $\nabla T=0.0105$.  
Figure~\ref{fig:ASP} shows the flow streamlines for the largest aspect
ratio here investigated, which can be compared with the streamlines of
a smaller aspect ratio in Fig.~\ref{fig:sch_flow}. Interestingly, the
flow patters are very similar, what verifies that the shape of
the flow streamlines is determined by the ratios $D/W$ and $H/D$. 

\begin{figure}
\hspace*{-0.5cm}
\includegraphics[width=0.5\textwidth]{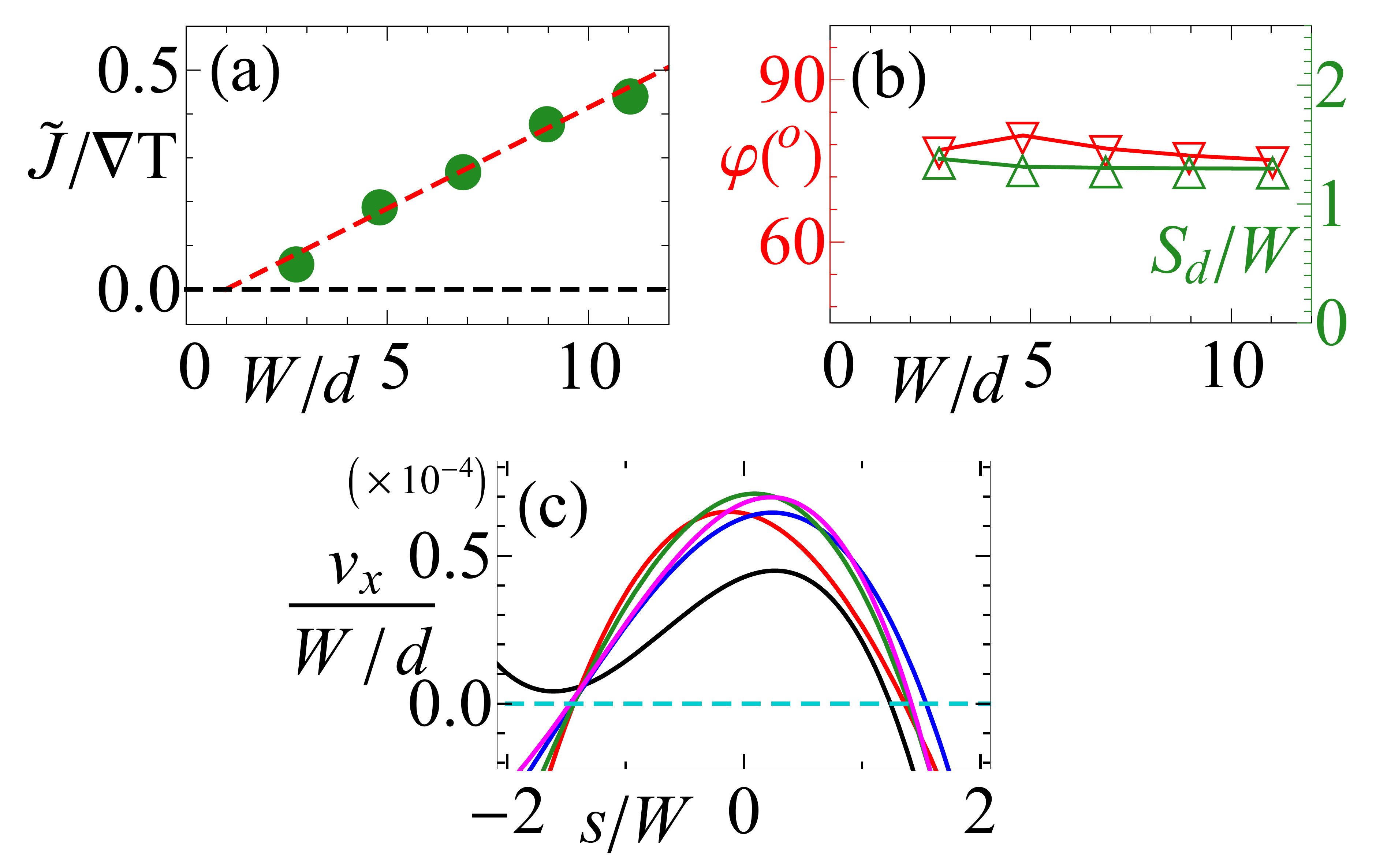}
  \caption{(a)~$\widetilde{J}/\nabla T$ as a function of aspect ratio
    $W/d$. (b)~Stagnation angle $\varphi$, and distance $S_d$ as a function of $W/d$. 
    (c)~Normalized velocity profile $v_x$ at the cross-section in between obstacles.}
	\label{fig:ASPf}
\end{figure}

The measured normalized averaged flux in Fig.~\ref{fig:ASPf}(a) very
nicely shows to linearly increase with $W/d$. Moreover, this
similarity can be inferred also from the constant stagnation angle
$\varphi$ as well as the rescaled stagnation distance $S_d/W$ for
different aspect ratios in Fig.~\ref{fig:ASPf}(b).  The flow velocity
profiles $v_x(z)$ perpendicular to the walls normalized by $W/d$ show
to collapse with a small deviation at the smallest aspect ratio,
Fig.~\ref{fig:ASPf}(c). This linear behavior implies that even with
small aspect ratio, we still obtain the same features of the study on
the flow patterns and the averaged flux $\widetilde{J}$.

\subsection{Mapping to physical units}

In order to provide an estimation of the actual pumping capability
under experimental conditions of the proposed microfluidic device, we
need to map the simulation units of the MPC model to those of real
physical systems. We employ a similar strategy as the one introduced
in Ref.~\cite{16ratchet}, where three relevant MPC quantities are
matched to real physical units. Very reasonable choices are to match
the average MPC temperature $T$ to $300$K, and the mass density of the
solvent to the density of water, $10^3 kg/m^3$. The choice for
  the length scale is however much more arbitrary, but we can make a
  realistic choice of channel widths between $20$ and $100\mu$m.  This
  allows us to identify $\sqrt{k_BT/m}$, and therefore the typical
  velocities in this work to be in the order of $5$ to $50~\mu$m/s,
  which are competitive for the design of microfluidic devices.
  Furthermore, higher velocities could be reached since
  Fig.~\ref{fig:ASPf} probes that the velocity linearly increases
  with the obstacle aspect ratio, given that the channel width to
  obtacle length ratio, $H/W$, and more importantly the temperature
  gradient, $\nabla T$, are kept constant.  On the other hand, it should be
 considered that the mesoscopic nature of the solvent makes it
impossible to simultaneously match all the relevant physical
quantities.  In this way, with the above parameter choices, we provide
a reasonable match to the ratio $\alpha_T\nabla T/\eta$, although not
to each of the involved quantities. And it is also very important that
most parameters employed in the simulations are chosen due to
computational efficiency, which is not related to any limitation of
the physical phenomenon. Thus, simulations of larger systems, with
smaller $\nabla T$, and larger $\alpha_T$ are in principle possible,
although not really meaningful at this stage. In this sense, the above
dimensions should be only taken as an educated guess, being other
sizes and velocities possible as well, in particular considering the
variaty of properties depending of the employed materials and system
conditions.

\section{Alternative setups}

\begin{figure}[h]
{\includegraphics[width=0.14\textwidth]{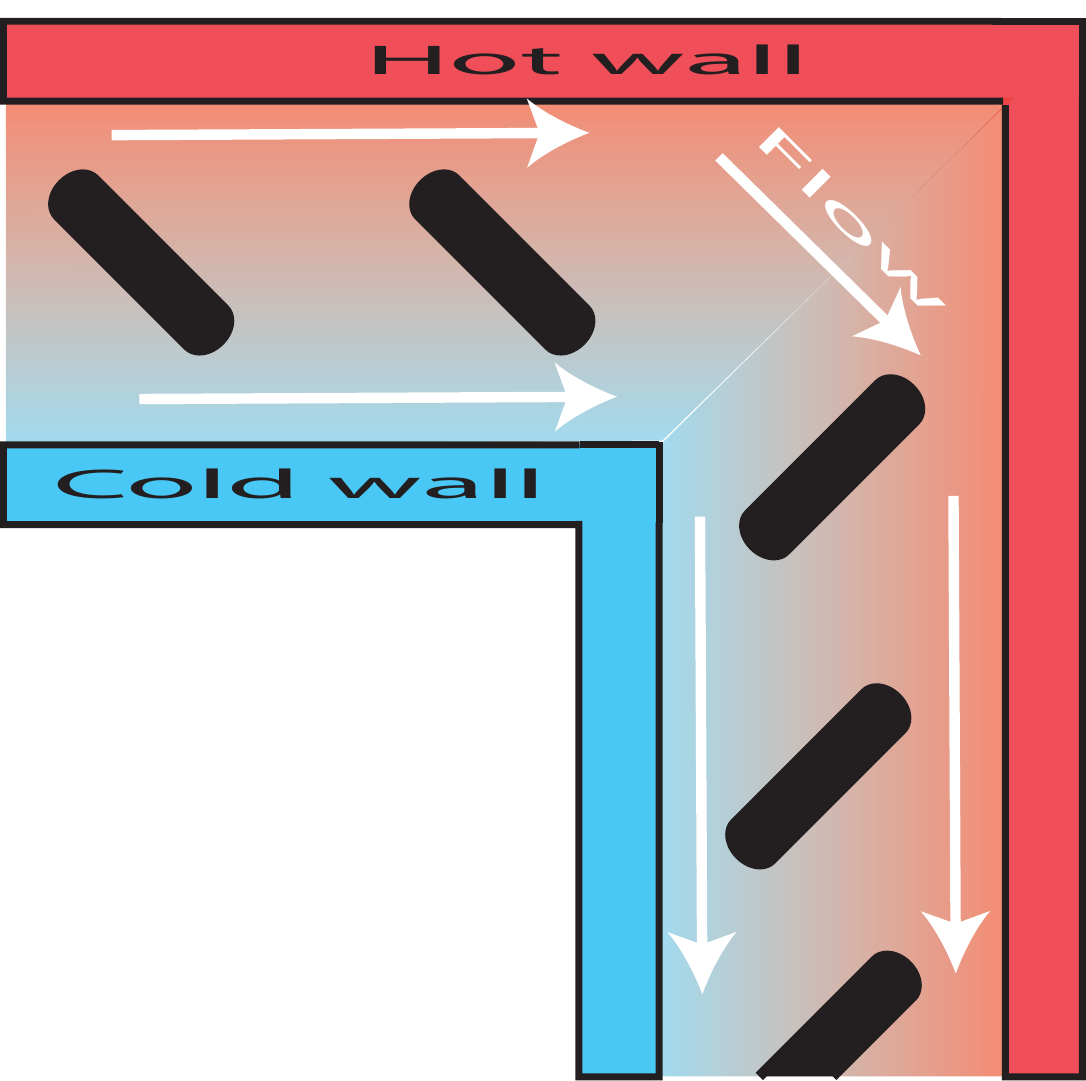}}
{\includegraphics[width=0.14\textwidth]{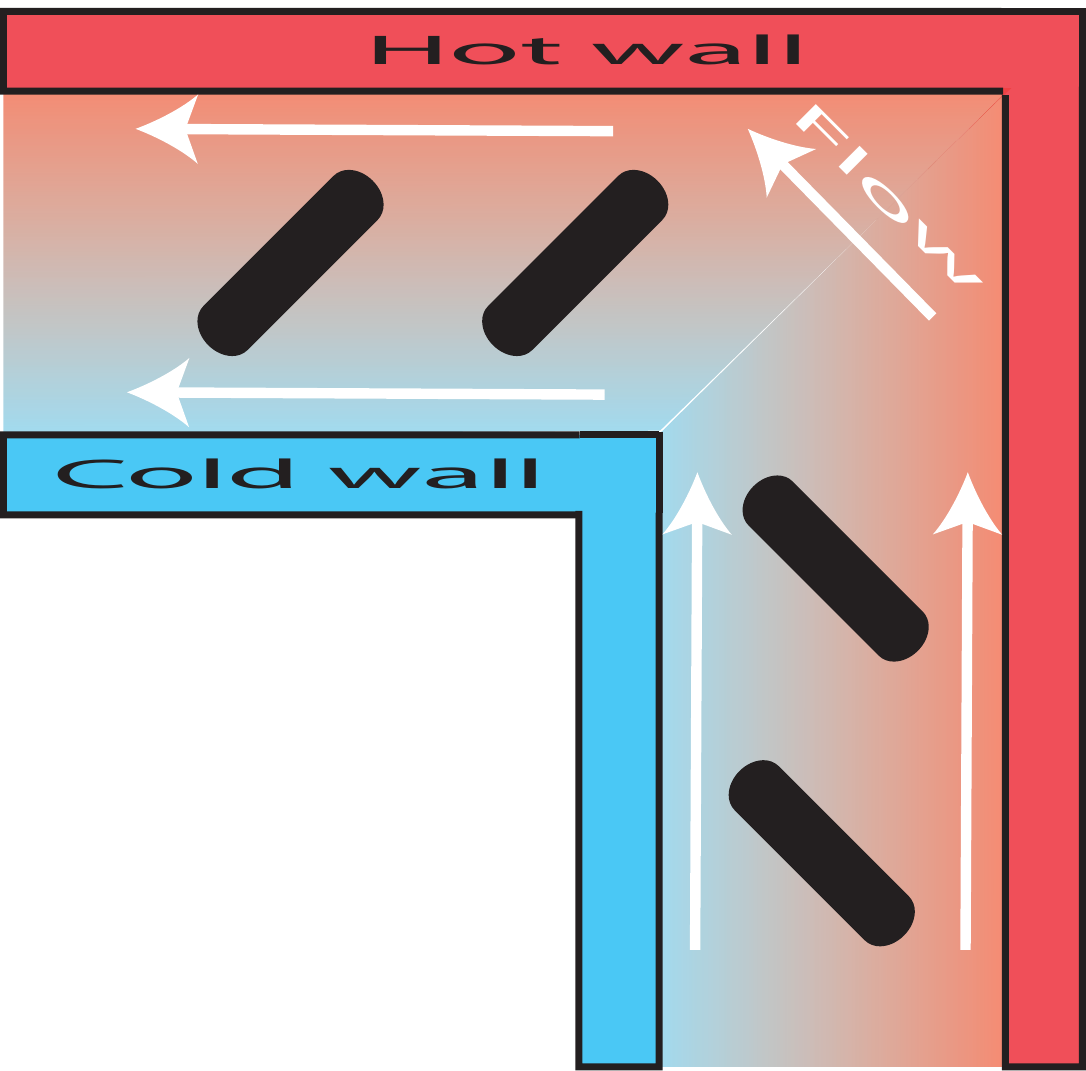}}
{\includegraphics[width=0.16\textwidth]{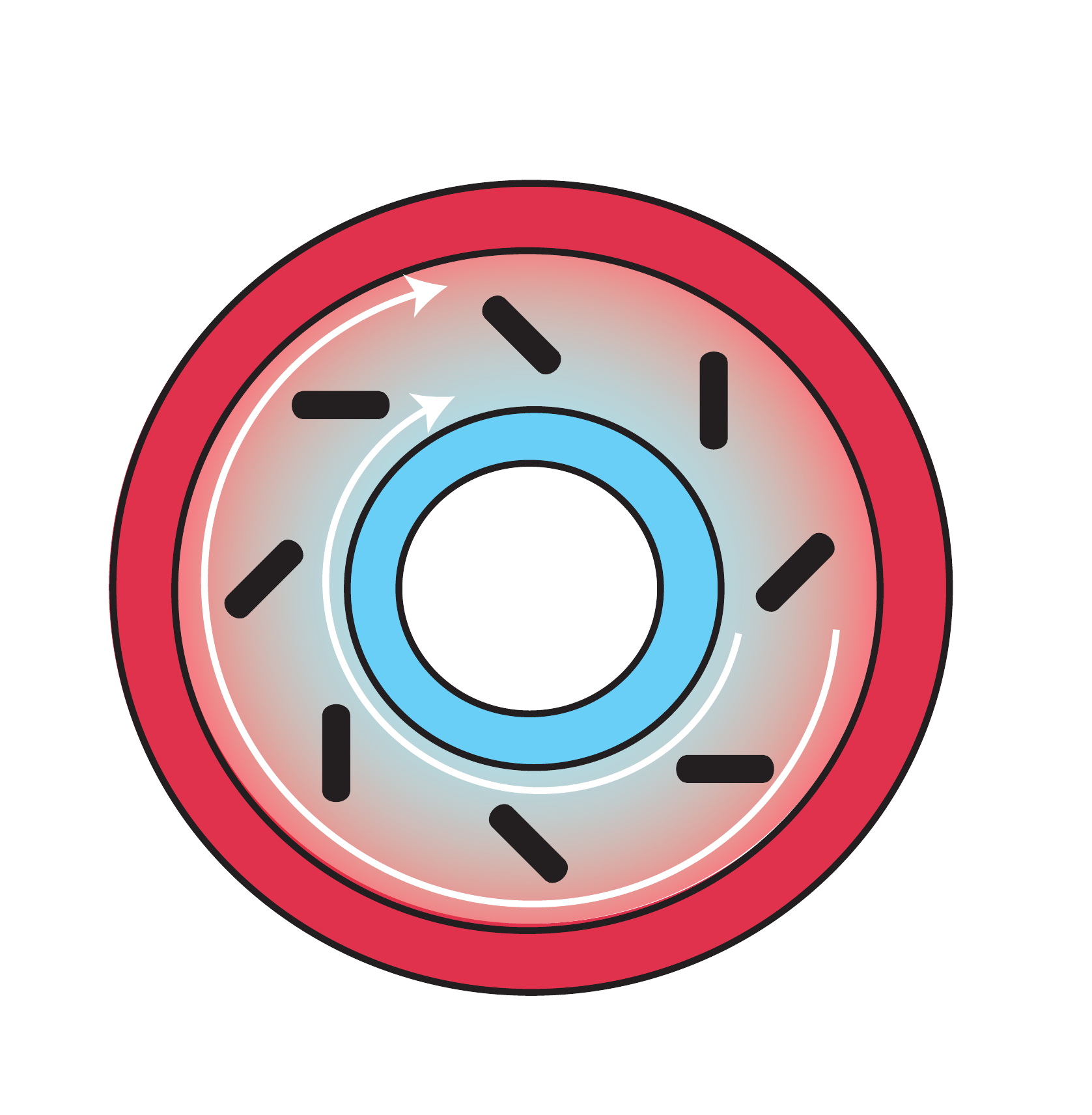}}
\caption{Sketches of non-straight channel configurations for
  anisotropic pumps, with the pillars kept at tilted orientations with
  respect to the walls. Note that although the flux direction
    here corresponds to $\chi_T > 0$, reverse direction is also
    possible by modifying the material properties.}
\label{fig:sketch1}
\end{figure}

The simulations presented in this work have so far considered only
straight microchannels with fixed obstacles placed equidistant from the
walls. These are though no restrictions for building functional
microfluidic devices based on the anisotropic thermophoretic effect.
In case that the pillars are not placed equidistant from the walls, the
fluid will stream with a non-symmetric pattern, but in general the net
fluid flux will remain. In the limiting situation in which the
obstacles are in contact with one of the walls, the system will be
similar in spirit to the phoretic ratcheted
microchannels~\cite{16ratchet,shen2016chemically}, where a net flux
with a shear-like velocity profile is generated. %%%
The channels do not need to be straight either, and corners or curved
geometries, as those shown in Fig.~\ref{fig:sketch1}, will not hinder
the fluid motion when the pillars are kept at the pertinent titled
orientation with respect to the walls. This is particularly important
for microfluidic applications of this phenomena, since it shows that
implementation in arbitrary geometries is possible.

Other interesting applications of the anisotropic phoretic effect in
microfluidics are fluid mixers and generators of alternating flow.
{\em Fluid mixers} can be obtained by building microchannels with elongated
pillars fixed in the middle of the channel with a fixed angle, but
with alternating positive and negative orientation, which would be
adopting a type of "w" or symmetric saw disposition. In this case a
type of Rayleigh-Bernard counter-rotating vortexes will be induced
with sizes determined by the obstacles positions and, rotating
velocities given by the temperature gradients and the thermophoretic
pillar properties. When two fluids converge into the phoretic
micropump with anisotropic obstacles, the mixing will be highly
enhanced due the intricate shape of the fluid flow, see for example
Fig.~\ref{figrugosity}(b) or Fig.~\ref{fig:Dflow}(b). For mixing
fluids~\cite{stone2004engineering,Alvaro15,Rallabandi15,Liu02}, the
pillars can, but do not need to be elongated, and no particular
orientation would be necessary.%%

\begin{figure}[h]
{\includegraphics[width=0.45\textwidth]{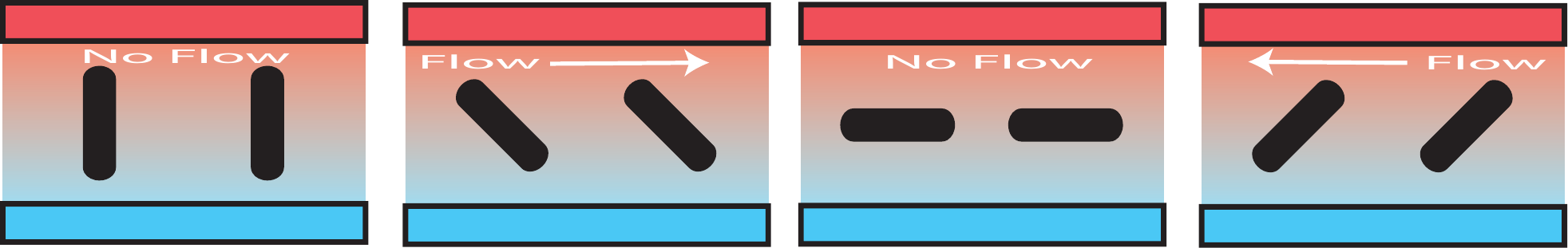}}
\caption{Sketch of the four consecutive configurations of a generator
  of alternating flow, in which states of no-flow, and flow with
  positive and negative directions can periodically alternate.}
	\label{fig:sketch2}
\end{figure}
The last structure that we discuss here is the {\em generators of
  alternating flow}, which emulates in some sense the alternating
current of electromagnetic devices. The idea is to change the
orientation of the pillars with time according to a prescribed
protocol. This could be realized if movable obstacles could be
engineered, or more realistically, by employing suspended elongated
particles with fixed centers of mass, and orientations externally
controlled by the use of for example laser tweezers, or rotating
magnetic fields. It is straightforward to predict that if the obstacle
orientation changes periodically with respect to the channel walls,
the net flux will change as indicated in Fig.~\ref{fig:sketch2}, from
being significant in one direction, then cancelling, then significant
in the opposite direction and then vanishing again.  An example of
systems that could be employed for this purpose are magnetizable
particles in a non-magnetic fluid, which in the presence of an
external rotating magnetic field will rotate with a constant
frequency~\cite{Melle2003}.  Although no simulation results are
presented for the structures discussed in this section, the extensive
results presented in the previous section serve as a strong
proof-of-concept.  This effect can be used not only as as a flow
alternator, but for example also as flow switch.

\section{Conclusions}
We have proposed a strategy to design microfluidic devices based on
anisotropic phoresis, this is the asymmetric response of elongated
objects to externally applied gradients, depending on their relative
orientation. 

The magnitude of the flow is determined by the phoretic properties of
the pillar surfaces and also by the channel geometric properties. We
have shown that this new design can facilitate the tunability of flow
velocity and pattern by solely altering the orientation, aspect ratio,
rugosity, or phoretic affinity (thermophobic, or thermophilic) of the
obstacles inside the same microchannel. The flux naturally increases
with the obstacle aspect ratio, and it is directly proportional to the
pillar phoretic anisotropic factor. This means that the flux direction
is not dictated by the thermophilic or thermophobic character of the
surface, but by the anisotropic thermophoretic factor which also
depends for example on the surface properties, such as
rugosity~\cite{ZT01}. This might be counter-intuitive, but it also
provides to the devices with an additional degree of versatility.
Interestingly, the current device uses anisotropic pillars at the
mid-channel, with an orientation which could eventually be manipulated
by for instance optical tweezers, this might inspire a new avenue of
microfluidic fabrication, such as fluidic switches, mixers, or flow
alternators. %%

The required temperature differences can be experimentally obtained in
two fundamentally different ways. One is by laser illumination when
one of the wall surfaces is metal
coated~\cite{Jiang2010,leonardo15gear}.  Such optical heating can be
flexibly and remotely controlled in a very precise and programmable
manner, and this microscale optical control can certainly find
applications in optofluidics~\cite{baigl12}. %%
Alternatively, the microchannel walls can be in contact with heat
reservoirs at different temperatures~\cite{braun08,vigolo2010}. This
contact heating can profit from existing residual heat flux, which
would eventually allow these device to harvest part of the waste
heat. Moreover, the proposed device can facilitate the cooling down of
microscale heat sources, such as microelectronic chips.
Furthermore, the current micropump needs only the
presence of the walls and a simple fluid, but it will also be
effective in the presence of a multicomponent fluid in a single phase
or in a multiphase situation, where the pillars could interact with
interfaces.

Although the micropumps here proposed have not yet been practically
realized, various existing experimental results can serve us as a {\em
  proof of concept}. An example is the osmotic flow which has shown to
be responsible for the formation of $2D$ thermophoretic colloidal
crystals close to a substrate~\cite{braun08,Leonardo2009}. Such flow
produces an inter-colloidal attraction, which maintains the colloidal
assembly only in the presence of a temperature gradient. A different
example can be found in the experimental results of a
self-thermocapillary asymmetric gear~\cite{leonardo15gear} showing
that a gear with an outer radius of $8~\mu$m can rotate with a maximum
angular velocity of $30$~rad/s when externally heated.

The discussion performed in this work has focused in the
thermophoretic case, but very importantly, all our arguments can be
straightforwardly generalized to other phoretic effects, such as {\em
  diffusiophoresis} or {\em electrophoresis}. This is for example the
case of a concentration gradient produced if one of the confining
walls would have a catalytic character. Simulation results have
already demonstrated anisotropic diffusiophoresis~\cite{YangMass}, as
well as the conceptual equivalence of, for example, the thermophoretic
gear~\cite{yang14gear}, and the catalytic
counterpart~\cite{Yang15cgear}. Based on this principle, a {\em proof
  of concept} has also been experimentally achieved for catalytic
self-electrophoretic microrotors~\cite{brooks19}, which show how platinum
microgears in solutions of hydrogen peroxide with an outer radius of
approximately $6~\mu$m can rotate with a maximum angular velocity of
$1.5$~rad/s. The mechanisms of the phoretic microgears and ratcheted
pumps~\cite{16ratchet,shen2016chemically} are different, but also closely
related with the basic principle discussed in this work, what strongly
supports the experimental feasibility of the here proposed devices in
both their thermal and catalytic versions.  %%
% The flexibility  of these micropumps and  is promising
% and can be attempted with different mechanisms and materials.

\section*{Acknowledgments}
M. R. thanks Andrea Costanzo for contributions in a very early stage
of this work.  The authors acknowledge financial support by China
Scholarship Council (CSC), and by the Bavarian Ministry of Economic
Affairs and Media, Energy and Technology within the joint projects in
the framework of the Helmholtz Institute Erlangen-N\"urnberg for
Renewable Energy (IEK-11) of Forschungszentrum J\"ulich. We also
gratefully acknowledge the computing time granted on the supercomputer
JURECA at J\"ulich Supercomputing Centre (JSC). M. Y. also
acknowledges financial support from the NSFC (No. 11674365). German
patent application (No. 102017003455.9) is pending for the work
described in this paper.

%\bibliographystyle{unsrt}
%\bibliography{ref}
%

\end{document}